\documentclass[11pt]{article}
\usepackage[round]{natbib}
\usepackage{a4wide,graphicx}
\usepackage{epstopdf}
\usepackage{lscape}
\usepackage[english]{babel}
\usepackage[latin1]{inputenc}
\usepackage{amssymb,amsmath,amsthm,latexsym,amsfonts,amscd,dsfont,enumerate,bbm}
\usepackage[pdftex]{color}
\usepackage[normalem]{ulem} 
\usepackage[]{hyperref}

\definecolor{mydarkgreen}{RGB}{34,169,34}
\definecolor{mydarkblue}{RGB}{72,61,169}

\hypersetup{
 colorlinks=true, 
 breaklinks=true, 
 urlcolor= mydarkblue, 
 linkcolor= mydarkblue, 
 citecolor= mydarkgreen,
 bookmarksopen=true, 
 pdfauthor={Callegaro, G, Fiorin, L., Pallavicini, A.}
 }

\newtheorem{theorem}{Theorem}[section]
\newtheorem{lemma}[theorem]{Lemma}
\newtheorem{proposition}[theorem]{Proposition}

\newtheorem{definition}[theorem]{Definition}

\newtheorem{remark}[theorem]{Remark}

\long\def\symbolfootnote[#1]#2{\begingroup%
\def\thefootnote{\fnsymbol{footnote}}\footnote[#1]{#2}\endgroup}

\title{Quantization goes Polynomial\thanks{\textbf{Acknowledgements:} this work was supported by the University of Padova research grant BIRD172407/17, ``New perspectives in stochastic methods for finance and energy markets''. Authors are grateful to Damien Ackerer, Damir Filipovic, Martino Grasselli, Martin Larsson and Daniele Marazzina for fruitful discussions.}}

\author{Giorgia Callegaro \thanks{Department of Mathematics ``Tullio Levi Civita'',
University of Padova, via Trieste 63, 35121 Padova, Italy. Email: gcallega@math.unipd.it ORCID ID: 0000-0001-9026-5261} \thanks{Corresponding author.}\\
\and
 Lucio Fiorin\thanks{Department of Mathematics ``Tullio Levi Civita'',
University of Padova, via Trieste 63, 35121 Padova, Italy. Email: luciofiorin@gmail.com ORCID ID: 0000-0002-0350-9473}
\and
Andrea Pallavicini \thanks{Department of Mathematics, Imperial College, London SW7 2AZ, UK and Banca IMI, Largo Mattioli 3, 20121 Milano, Italy. Email: a.pallavicini@imperial.ac.uk.
}}
\begin{document}

\maketitle

\begin{abstract}
Quantization algorithms have been successfully adopted to option pricing in finance thanks to the high convergence rate of the numerical approximation. In particular, very recently, recursive marginal quantization has been proven to be a flexible and versatile tool when applied to stochastic volatility processes. In this paper we apply for the first time quantization techniques to the family of polynomial processes, by exploiting their peculiar nature.  We focus our analysis on the stochastic volatility Jacobi process, by presenting two alternative quantization procedures: the first is a new discretization technique, whose foundation lies on the polynomial structure of the underlying process and which is suitable for vanilla option pricing, the second is based on recursive marginal quantization and it allows for pricing of (vanilla and) exotic derivatives. We prove theoretical results to assess the induced approximation errors, and we describe in numerical examples practical tools for fast vanilla and exotic option pricing.
\end{abstract}

\bigskip

\noindent {\bf AMS classification codes:} 65H35, 91G20, 91G60.

\section{Introduction}
Recently a new class of Markov processes, termed polynomial processes, has been introduced to model stock prices in view of financial applications. We refer to \cite{Cuchiero_poly} and \cite{Filipovic_poly} for an introduction and a review of the main properties of this family of processes, which includes, e.g., the Brownian motion, the geometric Brownian motion, Ornstein-Uhlenbeck processes, Jacobi processes, L\'evy processes and, more generally, affine processes. The main property of polynomial processes is that conditional expectations of polynomial functions of the process are again of polynomial type. Hereafter, we refer to this property as the ``polynomial property''. In particular, expected values of any polynomial of the process is again a polynomial in the initial value of the process, so that moments of all orders can be easily computed in closed form, up to a matrix exponential, even if the characteristic function of the process may be not known.

In this paper we focus on a particular polynomial process, the Stochastic Volatility Jacobi process (hereafter SVJ) first introduced in \cite{jacobi}, but our results can be extended to any polynomial model. The SVJ model is a diffusion model for stock prices where the log-price squared volatility follows a Jacobi process with values in some compact interval. It includes as limiting cases the Black-Scholes model and the Heston model, so that it can be viewed as a possible alternative to these models in practical applications of option pricing. 

The polynomial property allows to implement calibration algorithms for plain-vanilla options quoted by the market, which usually depend only on the marginal probability distribution of the underlying asset price at option expiry date. On the other hand, pricing exotic options also requires transition probabilities because of the path-dependent feature of such derivative products (think of time averages, continuous barriers and early redemptions). Joint probability distributions of the process at different times can be in principle derived by exploiting the polynomial property of the model, but the computational time rapidly explodes as the number of time observations increases. The work \cite{Filipovic_cubature} suggests one possible way to deal with this issue by introducing an approximating Markov chain via a moment matching condition. The space grid is made of the same set of points for all the discretization times. Since in general it is impossible to construct
a non-trivial Markov chain satisfying an exact $n$-th moment matching, with $n \ge 4$, the authors overcome this severe restriction by relaxing the exact moment matching and by looking for an approximation of the moments.

Working with a discrete state space clearly reduces the dimensionality of the problem and this is also what happens here with our alternative methodology, based on quantization. It is crucial to understand that our approach is different. More precisely, while the final target is the same, namely discretizing a stochastic process, the way this is done is totally different: quantization allows to optimally (in an $L^2$-sense) approximate the stochastic process at any time, so the output is an optimal grid which is not homogeneous with respect to time. We deem this is crucial when pricing path-dependent options over possibly long time horizons.

Quantization is a discretization technique for random variables, called vector quantization, and stochastic processes, known as functional quantization. The birth of quantization dates back to the 1950s, while its application in numerical probability and mathematical finance started in the 1990s. When applied to random vectors, quantization provides the best, according to a distance that is commonly measured using the Euclidean norm, possible approximation to the original distribution via a discrete random vector taking a finite number of values. Many numerical procedures have been studied to obtain optimal quadratic quantizers of random vectors even in high dimension and most of them are based on stochastic optimization algorithms, see \cite{Pag2014}
, which are typically very time consuming. Very recently vector quantization has been applied to recursively discretize stochastic processes. More precisely, the idea is approximating the random variables coming from the Euler scheme associated to a stochastic process, which is solution to a stochastic differential equation. Recursive marginal quantization, or fast quantization, introduced in \cite{PagSagna}, represents a new promising research field. Stationary quantizers of the stochastic process at fixed dates are obtained in a very fast and recursive way, to the point that recursive marginal quantization has been successfully applied to many models, starting from the classical ones and moving to local volatility models, such as in \cite{CalFioGra}, \cite{BCLP15}, \cite{Rudd17} and finally to stochastic volatility models, such as Stein and Stein, Stochastic Alpha Beta Rho, Heston and $\alpha-$hypergeometric, as done in \cite{CalFioGra2}, \cite{FiorinSagna}.

In the present paper we adopt two different quantization techniques to price plain-vanilla options and exotic derivative contracts within the SVJ model. The first technique is designed for plain-vanilla or European options, and it is based on the direct quantization of the price probability distribution obtained by exploiting the polynomial property. In this way we provide an alternative approach to the algorithm proposed in \cite{jacobi}. The second one is designed for exotic products, hence providing an alternative to \cite{Filipovic_cubature}, but it can be used also for European options, and it extends to multidimensional models the framework of \cite{CalFioGra2}, based on recursive marginal quantization. This second procedure does not rely on the polynomial property of the model, so that we can use it for path-dependent products without encountering dimensionality problems. Our analysis on SVJ quantization provides both practical tools to develop fast exotic option pricing algorithms and theoretical results to assess the approximation errors. A final Conclusion follows.

\medskip

The paper is organized as follows. In Section \ref{sec:tools} the SVJ model is presented. An introduction to quantization of random variables is given in Section \ref{recall}. Two different quantization approaches are then described: firstly, in Section \ref{sec:firstQ} quantization techniques are adapted to polynomial models, leading to new pricing formulas for plain-vanilla options, whose approximation error is discussed. Then, in Section \ref{sec:secondQ} recursive marginal quantization (which does not exploit the polynomial nature of our stochastic process) is introduced in a multidimensional setting and it is applied to price path-dependent exotic options. Numerical results for all the introduces algorithms, along with a discussion, are presented in Section \ref{sec:numerical}.

\section{The Stochastic Volatility Jacobi Model}
\label{sec:tools}

We consider a filtered probability space $(\Omega, \mathcal F,{(\mathcal F_t)}_{t \in [0,T]}, \mathbb Q)$, where $\mathbb Q$ is a risk neutral probability measure and where the filtration ${(\mathcal F_t)}_{t \in [0,T]}$ satisfies the usual hypotheses and models all the randomness in our model. We assume that the stock price process $S$ follows a SVJ model as in \cite{jacobi}, namely  we fix $0 \le v_{min} < v_{max}$ and we define
\begin{equation}
\label{eq:S_X}
S_t = e^{X_t}
\end{equation}%
where the dynamics of $(V,X)$ follows the stochastic volatility model
\begin{equation}\label{dynX}
\left\{
\begin{array}{rcl}
d V_t & = & \kappa (\theta  - V_t) dt + \sigma \sqrt{Q(V_t)} d W_{t} \\
d X_t & = & (r - \delta - V_t/2) dt + \rho \sqrt{Q(V_t)} dW_{t} + \sqrt{V_t - \rho^2 Q(V_t)} dW_{t}^{\perp}
\end{array}
\right.
\end{equation}%
with $X_0=x_0 \in \mathbb R, V_0=v_0 \in [v_{min} , v_{max}]$ and where the interest rate $r >0$, $\rho \in [-1,1]$, the mean reversion speed is $\kappa \ge 0$, the reversion level $\theta$ belongs to $[v_{min}, v_{max}]$,
\begin{equation} \label{Q}
Q(v) := \frac{(v - v_{min}) (v_{max} - v)}{ {( \sqrt{v_{max}} - \sqrt{v_{min}}  )}^2}
\end{equation}
and where $W$ and $W^{\perp}$ are independent standard Brownian motions. 
Notice that, recalling Equation \eqref{Q}, we have  $ v-Q(v) \ge 0$ with equality if and only if $v =\sqrt{v_{min} v_{max}}$ and $Q(v) \ge 0$ for every $v \in [v_{min}, v_{max}]$, so that the square roots in Equation \eqref{dynX} do not create any issue.
Clearly, we have that $\mathcal F_t = \mathcal F_t^{W} \vee \mathcal F_t^{W^{\perp}}$, $t \in [0,T]$. It is known that as special limiting cases of a SVJ model we obtain the Black-Scholes (take $v_0=\theta=v_{max}$) and the Heston model (take $v_{min}=0$ and $v_{max} \rightarrow \infty$).

\begin{remark} {\bf (Existence and Uniqueness of SVJ SDE Solution)}\\
The name SVJ is motivated by the model being clearly a stochastic volatility one, with the instantaneous squared volatility $V$ having a dynamics of Jacobi type, bounded on the interval $[v_{min}, v_{max}]$. Indeed, the following result holds (see \cite[Theorem 2.1]{jacobi}): for any deterministic initial state $(v_0,x_0) \in [v_{min}, v_{max}] \times \mathbb R$, there exists a unique solution $(V,X)$ to the system \eqref{dynX}, taking values in $[v_{min}, v_{max}] \times \mathbb R$. Furthermore, it is possible to show that if $(v_0,x_0) \in (v_{min}, v_{max}) \times \mathbb R$, then $(V_t,X_t)$ takes values in $(v_{min}, v_{max}) \times \mathbb R$ if and only if $\frac{\sigma^2 (v_{max} - v_{min})}{ ( \sqrt{v_{max}}  - \sqrt{v_{min}})^2} \le 2 \kappa \min \{  v_{max} - \theta , \theta - v_{min} \}  $.
\end{remark}

Moments in the SVJ model are known in closed form up to a matrix exponential. Indeed, if we write the generator $\cal G$ of the SVJ process, namely
$$
{\cal G} f(v,x) = b^\top(v)\nabla f(v,x) + \frac{1}{2} {\rm Tr}\left( a(v) \nabla^2 f(v,x) \right)
$$%
with drift vector $b(v)$ and the diffusion matrix $a(v)$ given by
$$
b(v) =
\begin{bmatrix}
\kappa (\theta - v) \\ r - \delta - v/2
\end{bmatrix}
\;,\quad
a(v) =
\begin{bmatrix}
\sigma^2 Q(v) & \rho\sigma Q(v) \\ \rho\sigma Q(v) & v
\end{bmatrix}
$$%
we have that $\cal G$ maps any polynomial of degree $n$ onto a polynomial of degree $n$ or less as shown in \cite{Filipovic_poly}. As a consequence it is possible to evaluate the conditional moments of $(V_T,X_T)$ as follows. Let ${\rm Pol}_n$ be the vector space of polynomials in $(v,x)$ of degree less than or equal to $n$. For any positive integer $\mathsf N$, we term $\mathsf M = (\mathsf N + 2)(\mathsf N + 1)/2$ the dimension of ${\rm Pol}_{\mathsf N}$, we introduce a basis ${h_1(v,x),\ldots,h_{\mathsf M}(v,x)}$ of polynomials of ${\rm Pol}_{\mathsf N}$, and we denote by $G$ the matrix representation of the linear map $\cal G$ restricted to ${\rm Pol}_{\mathsf N}$ with respect to this basis. Thus, from  Theorem 3.1 in \cite{Filipovic_poly} we get that for any polynomial $p\in{\rm Pol}_{\mathsf N}$ we have
$$
\mathbb{E} \left[ p(V_T,X_T) | {\cal F}_t \right] =
\begin{bmatrix}
h_1(V_t,X_t) & \ldots & h_{\mathsf M}(V_t,X_t)
\end{bmatrix}
e^{(T-t)G}{\vec p}
\quad
\forall t<T
$$%
where ${\vec p} \in \mathbb{R}^{\mathsf M}$ is the coordinate representation of the polynomial $p(v,x)$ with respect to the basis $h$. In this paper we term this relationship as \textit{polynomial property}.

We recall here the technical results on the closed form pricing of European options, that we will need from now on. Let us define the weighted Lebesgue space
$$
L_{w}^{2}=\bigg\{ f  \ \textrm{measurable} : ||f||_{w}^{2} = \int_{\mathbb{R}} f^{2}(x) w(x) dx < \infty \bigg\},
$$%
equipped with the scalar product
$$
\langle f, g \rangle_{w} = \int_{\mathbb{R}} f(x) g(x) w(x) dx,
$$%
where $w$ is the Gaussian weight  function, i.e., the Gaussian density with mean $\mu_{w}$ and variance $\sigma_{w}^{2}$. The Hilbert space $L^{2}_{w}$ admits an orthonormal basis of  generalized Hermite polynomials $H_{n}$, $n \ge 0$ (notice that the degree of $H_n$ is equal to $n$), given by
\begin{equation}\label{genherm}
H_{n}(x) = \frac{1}{\sqrt{n !}} \mathcal{H}_{n} \left( \frac{x - \mu_{w}}{\sigma_{w}} \right),
\end{equation}%
where $\mathcal{H}_{n}$ are the probabilist Hermite polynomials defined as
\begin{equation}\label{herm}
\mathcal{H}_{n}(x) = (-1)^{n} e^{\frac{x^{2}}{2}} \frac{d^{n}}{d x^{n}} e^{- \frac{x^{2}}{2}}.
\end{equation}%
If we assume that $g_{T}$ is the density of the log price $X_{T}$, then we can define $\ell(x) = \frac{g_{T}(x)}{w(x)}$. 
\begin{remark} \label{rem:ell_in_L^2} {\bf (Assumptions on SVJ Parameters)}\\
We need $\ell \in L_w^2$, so that from now on, applying Corollary 3.3 in \cite{jacobi}, we assume that
$$
v_{min} > 0, \quad \rho^2 <1, \quad \sigma_w^2 > \frac{v_{max} T}{2}.
$$%
\end{remark}
Our aim being pricing a European option with payoff $f \in L^{2}_{w}$ (notice that both Call and Put options have payoffs belonging to $L^2_w$), we get
$$
\mathbb{E}[f(X_{T})]=\int_{\mathbb{R}} f(x) g_{T}(x) dx = \int_{\mathbb{R}} f(x) \ell(x) w(x) dx = \langle f, \ell \rangle_{w}.
$$%

Since $L^{2}_{w}$ is an Hilbert space with orthonormal basis $H_{n}$ defined in Equation \eqref{genherm}, we can rewrite the previous formula as
\begin{equation}\label{exppol}
\mathbb{E}[f(X_{T})] = \sum_{n \geq 0} f_{n} \ell_{n},
\end{equation}%
for the \textit{Fourier coefficients}
\begin{equation}\label{fourier}
f_{n}=\langle f, H_{n} \rangle_{w},
\end{equation}%
and the \textit{Hermite moments}
\begin{equation}\label{ell_n}
\ell_{n}=\langle \ell, H_{n} \rangle_{w} = \int_{\mathbb{R}} H_{n}(x) \ell(x) w(x) dx = \int_{\mathbb{R}} H_{n}(x) g_{T}(x) dx.
\end{equation}%

Notice that the last equality shows that $\ell_{n}$ is a linear combination of moments of $X_{T}$, since $H_{n}$ is a polynomial. It is then possible to compute $\ell_{n}$ in closed form, because of the polynomial nature of the process.

\section{Essentials on Quantization}
\label{recall}

In this section we introduce optimal quadratic quantization of a random variable (also known as vector quantization), which will be a necessary tool toward a specific discretization of a stochastic process, known as recursive marginal quantization (henceforth RMQ). We refer to \cite{Graf:book} and \cite{Pag2014} for vector quantization and to \cite{PagSagna} for the first paper on RMQ\footnote{We also have to refer to the website: \url{http://www.quantize.maths-fi.com}, where grids of the $d$-dimensional Gaussian distributions $\mathcal N(0; I_d)$, for $N = 1$ up to $10^4$ and for $d = 1, \dots, 10$ can be downloaded.}.

Optimal quadratic quantization answers the following question: is it possible (and how) to optimally approximate, in an $L^2$-sense, a continuous random variable $X$ by a discrete one, $\widehat X$, taking a finite number of values?

The interest in such a discretization $\widehat X$ is clear: expectations in the form $\mathbb E [h(X)]$ (for sufficiently regular functions $h$) would be approximated by finite sums. Let us be more precise. We consider a real valued random variable $X$ defined on $(\Omega,\mathcal{F},\mathbb{P})$, having probability distribution $\mathbb P_X$ and admitting a finite second order moment. A quantization grid of level $N, N \ge 1$, is a subset  of $\mathbb R$, $\Gamma=\{ x_1, \dots, x_N\}$ (here $N$ will be fixed, so that for simplicity we drop the dependence on $N$ in the notation), of size at most $N$ having pairwise distinct components. A quantization function, or \emph{quantizer}, is a $\Gamma$-valued Borel function $q: \mathbb R \rightarrow \Gamma$ and quantizing $X$ means projecting $X$ on $\Gamma$ following the closest neighbor rule
\begin{equation}
q(X) = {\rm Proj}_{\Gamma}(X) := \sum_{i=1}^N x_i 1\!\!1_{C_i(\Gamma)}(X)
\end{equation}%
where ${(C_i(\Gamma))}_{1 \le i \le N}$ is a Borel partition of $(\mathbb R, \mathcal B(\mathbb R))$, also known as Voronoi partition, satisfying $ C_i(\Gamma) \subset \left\{  \xi \in \mathbb R : \vert \xi - x_i \vert = \min_{i \le j \le N} \vert \xi - x_j \vert  \right\}, \quad i = 1 \dots, N $. We will use the notation $\widehat X^{\Gamma}$ or $\widehat X$  (when no ambiguity is possible with respect to the grid) to denote the Voronoi $\Gamma$-quantization of $X$:
$
\widehat X^{\Gamma} = \widehat X=  q(X). 
$%

The $L^2$-error coming from such a discretization is given by
$$
e_{N}(X,\Gamma) := \Vert X - q(X) \Vert_2  =  \Vert \min_{1 \leq i \leq N} \vert X - x_i \vert \Vert_2
$$%
where ${||X||}_2 := {\left[\mathbb E({|X|}^2) \right]}^{1/2}$ is the usual $L^2$-norm and the aim of optimal quadratic quantization is finding a grid $\Gamma$, with size at most $N$, which minimizes the \emph{distortion function} defined below (see \cite[Equation (3.4)]{Graf:book}).
\begin{definition}
\label{defDistortion}
Let $X$ be a real valued random variable belonging to $L^2(\mathbb P)$. The $L^2$-distortion function is a positive valued function defined on ${\mathbb R}^N$ by
\begin{equation}\label{distortion}
D:  (x_1, x_2, \dots, x_N) \longmapsto \mathbb E \big[  \min_{1 \le i \le N} \vert X - x_i \vert ^2 \big]  = e_{N}(X,\Gamma)^2.
\end{equation}%
\end{definition}%

Concerning the existence and uniqueness of an optimal grid, it is possible to show, see e.g. \cite[Prop. 1.1]{Pag2014}, that if $X \in L^2(\mathbb P)$, then the distortion function $D$ attains (at least) one minimum $\Gamma^{\star}$. The grid $\Gamma^{\star}$ and $ {\rm Proj}_{\Gamma^{\star}}$ are called \emph{optimal quadratic quantizers}, respectively. In the case when card(supp$(\mathbb P_X)) \ge N$, then $\Gamma^{\star}$ has pairwise distinct components. Moreover $\lim_{N \rightarrow +\infty} e_{N}(X) =0$ and the convergence rate is given by the well-known Zador theorem (see \cite{Graf:book})
\begin{equation}
\label{eq:zador}
\min_{\Gamma,\ \vert \Gamma \vert = N} \Vert X - q(X) \Vert_2 =  \min_{\Gamma,\ \vert \Gamma \vert = N} e_{N}(X,\Gamma)  = Q_2(\mathbb{P}_X) N^{-1} + o \big(N^{-1} \big) 
\end{equation}%
where $Q_2(\mathbb P_X)$ is a nonnegative constant. Then, it is natural to approximate an expected value of the form $\mathbb{E}\left[h(X) \right]$ in the following way:
\begin{equation}\label{expval}
\mathbb{E}\left[h(X) \right] \approx \mathbb{E}\left[h(\widehat X) \right] = \sum_{i=1}^{N} h(x_{i}) \mathbb{P} \left( \widehat X = x_{i} \right) = \sum_{i=1}^{N} h(x_{i}) \mathbb{P} \left(  X \in C_{i}(\Gamma) \right).
\end{equation}%
Moreover, as mentioned in \cite[Theorem. 5.1]{Graf:book}, as soon as $\mathbb P_X$ is absolutely continuous with respect to a log-concave density, then there exists exactly one optimal quantization grid of level $N$.

\begin{remark} {\bf (Quantization vs.\ Monte Carlo Pricing Error)}\\
When computing $\mathbb{E}\left[h(X) \right]$, where $X$ is the value at maturity of an underlying asset and $h$ is a Lipschitz (payoff) function with Lipschitz constant $ [h]_{\text{Lip}}$, then, by applying Jensen's inequality,
$$
\big \vert \mathbb{E} [h(X)]  - \mathbb{E}[h(\widehat X^{\Gamma})] \big \vert \leq [h]_{\text{Lip}} e_{N}{(X,\Gamma)}.
$$%
In particular, the error coming from pricing via quantization decays at the rate $\frac{1}{N}$, as opposed to the Monte Carlo error, ruled by the Central Limit Theorem, which is of order $\frac{1}{\sqrt{N}}$.
\end{remark}

The last crucial point is how to obtain an optimal quantizer. It is known that the distortion function is differentiable at  any $N$-tuple having pairwise distinct components $\Gamma = \{ x_1,\dots,x_N\}$ (see \cite[Lemma 4.10]{Graf:book} or \cite[Prop. 1.1]{Pag2014}), with differential
\begin{equation}\label{nablaD}
\nabla D(x_1,\dots,x_N) = 2 {\left( \int_{C_i(\Gamma)} (x_i - \xi) d \mathbb P_X (\xi)   \right)}_{1 \le i \le N} 
=  2 {\left( \mathbb E \big[ 1\!\!1_{X \in C_i(\Gamma)} (x_i - X) \big]   \right)}_{1 \le i \le N},
\end{equation}%
so that many stochastic algorithms looking for zeros of the gradient of $D$ have been developed. These include gradient descent and fixed point procedures and we refer to \cite[Section 3]{Pag2014} for a detailed overview. Critical points of the distortion function are called \textit{stationary quantizers}. Optimal quantizers are stationary, but the viceversa is not true in general. Of course, stationary quantizers are in general not unique. From \eqref{nablaD}, the determination of stationary quantizers boils down to finding the solution of 
\begin{equation}
\label{masterEq}
\mathbb E \big[ 1\!\!1_{X \in C_i(\Gamma)}  X  \big] - x_{i} \mathbb{P} \left(X \in C_i(\Gamma)\right) = 0 \quad  \forall i \in \{ 1, \dots, N\}
\end{equation}%
known as \textit{Master Equation} (it is, indeed, a system of $N$ equations in $N$ unknowns $x_1,\dots, x_N$). Moreover, when the gradient itself is differentiable, it is possible to apply the classical Newton-Raphson procedure.

If the density of $X$ is known, then it is possible to explicitly write the system in \eqref{masterEq} in closed form and to find its solution. On the other hand, when $X$ is the asset price at maturity, typically the density of the process is not explicitly known, except in trivial cases, and finding stationary quantizers becomes numerically interesting.

We conclude by sketching the ideas behind RMQ, which allows to use vector quantization to recursively discretize a stochastic process. More precisely, \cite{PagSagna} introduced RMQ to discretize a stochastic process $Y$ (in dimension one) by recursively working on the random variables $\{( Y_{t_k}) \}_{k =0, \dots, M}$ associated with the Euler-Maruyama scheme. So, starting from a time discretization of a stochastic process on $\{ {t_k} \}_{k =0, \dots, M}$, vector quantization provides a state space discretization of every random variable $Y_{t_k}$. The essence of RMQ lies on the knowledge of the conditional law of $(Y_{t_k}\vert Y_{t_{k-1}}), k=1,\dots, M$, which allows to recursively quantize the marginals $\{( Y_{t_k}) \}_{k =0, \dots, M}$ via a Newton-Raphson procedure (the gradient and the Hessian of the distortion function are explicit).

In the following Section \ref{sec:secondQ} we provide more details on RMQ applied to our multidimensional setting.

\section{Quantization of a Polynomial Process}
\label{sec:firstQ}

We will now discuss how to deal with polynomial processes. In particular, we consider the SVJ model, even if our approach is general and flexible enough to be applied to any polynomial process. We only consider the SVJ model since it is a good representative for the class of diffusive polynomial processes. 
As a first approach, in this section, we exploit the polynomial property and we focus on the quantization of the log price process $X$ at a fixed date. Then, in Section \ref{sec:secondQ}, in order to deal with path-dependent options, we forget about the polynomial nature of $(V,X)$ and we extend the general framework in \cite{CalFioGra2} to discretize the bidimensional process $(V,X)$ at a whole set of dates via RMQ.

\subsection{Exploiting the Polynomial Property}
\label{sec:poly}

In this section we consider the problem of finding a stationary quantizer of the log price process $X$ at a given time $T$. 

%
%
The main result of this section is the possibility of writing the Master equation \eqref{masterEq} in closed form, thanks to the polynomial nature of our processes.

\begin{theorem}
\label{thm:masterEq}
{\bf (Polynomial Process Quantization)}\\
Consider the Master equation 
\begin{equation}\label{masterSys}
\mathbb{E} \left[ \left( X_{T} - x_{i} \right) 1\!\!1_{X_{T} \in C_{i}(\Gamma)} \right] = 0, \quad i=1, \dots, N
\end{equation}%
and its $i$-th component
\begin{equation}\label{masterSys_i}
E_{i}(x_{1}, \dots, x_{N}):= \  \mathbb{E} \left[ \left( X_{T} - x_{i} \right) 1\!\!1_{X_{T} \in C_{i}(\Gamma)} \right]  = 0 
\end{equation}%
where $\displaystyle C_{i}(\Gamma) = \left[ \frac{x_{i-1} + x_{i}}{2}, \frac{x_{i} + x_{i+1}}{2} \right] $,  $C_{1}(\Gamma) = \left[- \infty, \frac{x_{1} + x_{2}}{2} \right]$ and $C_{N}(\Gamma) = \left[\frac{x_{N-1} + x_{N}}{2} , +\infty \right]$. In our setting Equation \eqref{masterSys_i} reads
\begin{equation}\label{masterpol}
\sum_{n \geq 0} f^{i}_{n} \ell_{n} = 0,
\end{equation}%
where $\ell_{n}$ are the Hermite moments defined in \eqref{ell_n} and where the (Fourier) coefficients $f^{i}_{n}$ are given by
$$
f^{i}_{n}=h_{n}\left(  \frac{x_{i-1} + x_{i}}{2} \right) - h_{n}\left(  \frac{x_{i} + x_{i+1}}{2} \right) - x_{i} \left(  l_{n}\left(  \frac{x_{i-1} + x_{i}}{2} \right) - l_{n}\left(  \frac{x_{i} + x_{i+1}}{2} \right)   \right)
$$%
with
\begin{align}\label{coeffh}
h_{0}(K) & = \sigma_{w} \phi\left( \frac{K - \mu_{w}}{\sigma_{w}} \right) + \mu_{w} \Phi \left( \frac{\mu_{w}-K}{\sigma_{w}}\right) \nonumber \\
h_{1}(K) &  = \sigma_{w} \left[ \frac{K - \mu_{w}}{\sigma_{w}} \phi \left(\frac{K - \mu_{w}}{\sigma_{w}} \right)  + \Phi \left( \frac{K - \mu_{w}}{\sigma_{w}} \right) \right] + \mu_{w} \phi \left(\frac{K - \mu_{w}}{\sigma_{w}} \right) \nonumber \\
h_{n}(K) &  = \frac{1}{\sqrt{n!}} \phi\left( \frac{K - \mu_{w}}{\sigma_{w}} \right) \bigg[ \sigma_{w} \mathcal{H}_{n}\left( \frac{K - \mu_{w}}{\sigma_{w}} \right) +n \sigma_{w} \mathcal{H}_{n-2}\left( \frac{K - \mu_{w}}{\sigma_{w}}\right) \\
& + \mu_{w} \mathcal{H}_{n-1}\left( \frac{K - \mu_{w}}{\sigma_{w}} \right)\bigg], n \geq 2, \nonumber
\end{align}%
and
\begin{align}\label{coeffl}
l_{0}(K)&= \Phi \left( \frac{K - \mu_{w}}{\sigma_{w}} \right) \nonumber\\
l_{n}(K)& = \frac{1}{\sqrt{n !}} \mathcal{H}_{n-1} \left( \frac{K - \mu_{w}}{\sigma_{w}} \right) \phi \left( \frac{K - \mu_{w}}{\sigma_{w}} \right), n \geq 1,
\end{align}%
and where $\phi$ and $\Phi$ are, respectively, the density and the cumulative distribution functions of a standard univariate Gaussian random variable.
\end{theorem}

\begin{proof}
See Appendix \ref{appA}.
\end{proof}

\begin{remark}
The results of Theorem \ref{thm:masterEq} are general, in that they can be applied to other polynomial models, by suitably modifying the cofficients $\ell_n$.
\end{remark}

\subsection{Calculation of a Stationary Quantizer}
\label{sec:subopt}

Even if Equation \eqref{masterSys_i} can be written in closed form for every $i=1, \dots, N$, it is impossible to find an analytical expression for the solution to the nonlinear system, which corresponds to the stationary quantizer. Hence, we need to solve this system numerically.  As already noted in Section \ref{recall}, the literature suggests the Newton-Raphson method as the best first choice to tackle the system of equations. The proposition below provides the Jacobian matrix to be used in the Newton-Raphson procedure.

\begin{proposition}
\label{propJacobian}
Consider the system of equations \eqref{masterSys}
\begin{equation}
\left\{
\begin{array}{rcl}
E_{1}(x_{1}, \dots, x_{N}) &:=& \mathbb{E} \left[ \left( X_{T} - x_{1} \right) 1\!\!1_{X_{T} \in C_{1}(\Gamma)} \right]  = 0 \\
E_{2}(x_{1}, \dots, x_{N}) &:=& \mathbb{E} \left[ \left( X_{T} - x_{2} \right) 1\!\!1_{X_{T} \in C_{2}(\Gamma)} \right]  = 0 \\
& \vdots &\\
 E_{N}(x_{1}, \dots, x_{N}) &:=& \mathbb{E} \left[ \left( X_{T} - x_{N} \right) 1\!\!1_{X_{T} \in C_{N}(\Gamma)} \right]  = 0 .
\end{array}
\right.
\end{equation}%
When $X$ is a polynomial process with Hermite moments $\ell_{n}$, the Jacobian matrix $J$ of the vector function $E = ( E_{1}, \dots, E_{N} )$ is tridiagonal and symmetric, and its components have the following form:
\begin{align}\label{jacobian}
J_{i,i-1} & = \frac{1}{2} \left( \frac{x_{i} - x_{i-1}}{2} \right) g_{T} \left( \frac{x_{i-1} + x_{i}}{2} \right) \quad i=2, \dots, N \nonumber \\
J_{i,i} & =  J_{i,i-1} + J_{i,i+1} - \mathbb{P} \left( X_{T} \in C_{i} (\Gamma) \right) \quad i=1, \dots, N,\nonumber
\end{align}%
with $J_{1,0}=J_{N,N+1}=0$ and where
\begin{equation}\label{density}
g_{T}(x) = \sum_{n \geq 0} \ell_{n} H_{n}(x) w(x),
\end{equation}%
\begin{equation}\label{weights}
\mathbb{P} \left( X_{T} \in C_{i} (\Gamma) \right) = \sum_{n \geq 0} \ell_{n} \left(  l_{n}\left(  \frac{x_{i-1} + x_{i}}{2} \right) - l_{n}\left(  \frac{x_{i} + x_{i+1}}{2} \right)   \right),
\end{equation}%
and the coefficients $l_{n}$ are computed in \eqref{coeffl}. 
\end{proposition}

\begin{proof}
See Appendix \ref{appA}.
\end{proof}

The Newton-Raphson algorithm has then the following structure: starting from an initial guess $\Gamma^{(0)}$, the $k$-th iteration is
\begin{equation}
\Gamma^{(k)} = \Gamma^{(k-1)} - J^{-1}\left(\Gamma^{(k-1)} \right) E \left(\Gamma^{(k-1)} \right) \quad k=1, 2,  \dots
\end{equation}%
where $E=(E_1, \dots, E_N)$ is defined in Proposition \ref{propJacobian} and it is computed thanks to Theorem \ref{thm:masterEq}. Given a stopping criterion, the final iteration gives the stationary quantization grid $\Gamma^{*}$.

Let us assume now that we have found the solution $\Gamma^{*}=\{x_{1}^{*}, \dots, x_{N}^{*} \}$, which is the stationary quantization grid associated to $X_{T}$. In order to compute an expected value as in \eqref{expval}, we need to know the weights associated to every Voronoi cell $C_{i}(\Gamma^{*})$, for $i=1, \dots, N$. The weights are straightforwardly given by \eqref{weights}.

\subsection{Analysis of the Approximation Error}
\label{sec:err}

We focus on pricing of at time $0$ of a European option with payoff $f$. We consider, without loss of generality, a Call option written on $S$ having expiry $T >0$ and strike price $K$,  i.e.,
$$
f = f(S_T) = {(S_T - K)}^+.
$$%
Of course, as mentioned before, the results in this section are also valid for Put options.\\
In what follows we will need the following three versions of the price:
\begin{itemize}
\item $\pi_{f}$ is the exact price at time $0$, i.e.,
$$
\pi_{f} := 
 \mathbb E^{\mathbb Q} \left[ e^{- r T} {(S_T - K)}^+ \right] =  e^{- r T } \int_{\mathbb{R}} \left( e^{x} -  K\right)^{+} g_{T} (x) dx,
$$%
where $g_{T}$ is the density of the log price $X$ at time $T$, given by \eqref{density}. This formula contains an infinite sum, so the function $g_T$ function is not computable in closed form.
\item $\pi_{f}^{(M)}$ is the price computed using the polynomial approximation at level $M$, i.e., approximating the density 
$g_{T}(x)$ with 
\begin{equation}\label{g_T^(M)}
g_{T}^{(M)}(x) = \sum_{n = 0}^{M} \ell_{n} {H}_{n}(x) w(x),
\end{equation}%
namely
$$
\pi_{f}^{(M)} = e^{- r T } \int_{\mathbb{R}} \left( e^{x} -  K\right)^{+} g_{T}^{(M)} (x) dx = \sum_{n=0}^{M} \ell_{n} f_{n},
$$%
\item $\widehat{\pi}_{f}^{(M,N)}$ is the price computed by approximating the log-spot price at maturity by means of quantization on a grid with $N$ points:
\begin{eqnarray}
\widehat{\pi}_{f}^{(M,N)}
&=& e^{- r T } \sum_{i=1}^N \int_{C_{i}(\Gamma^X )} {(e^{x_i} - K)}^+ g_T^{(M)}(x) dx \nonumber\\
&=& e^{- r T } \sum_{i=1}^N  {(e^{x_i} - K)}^+ \mathbb P \left(  X_T^{(M)} \in  C_i(\Gamma^X)  \right)
\end{eqnarray}%
where  $\Gamma^X = \{x_{1}, \dots, x_{N} \}$ is the optimal quantizer relative to $X_T^{(M)}$, the log price with (approximate) density $g_T^{(M)}$. We denote by $\widehat{X}_T^{(M,N)}$ the quantization of $X_T^{(M)}$. Notice that we also have
$$
 \widehat{\pi}_{f}^{(M,N)} = e^{-r T} \sum_{i=1}^{N} \left( e^{ x_{i}} - K\right)^{+} \mathbb{P}\left(  \widehat{X}_T^{(M,N)} = x_{i} \right).
$$%
\end{itemize}

The accuracy of our methodology is studied by analyzing the (asymptotic) behavior of the price approximation, namely
\begin{equation}
\label{quanterror}
\textrm{err}_{M,N}:= \left|  \pi_{f} - \widehat{\pi}_{f}^{(M,N)} \right|.
\end{equation}%

We split the error in two parts that we study separately:
$$
\textrm{err}_{M,N} \leq  \left|  \pi_{f} - \pi_{f}^{(M)} \right| +  \left|  \pi_{f}^{(M)} - \widehat{\pi}_{f}^{(M,N)} \right| = \textrm{err}^1_{M} + \textrm{err}^2_{M,N},
$$%
where $\textrm{err}^1_{M} := \left|  \pi_{f} - \pi_{f}^{(M)} \right| $ is the truncation error $|\varepsilon^{(M)}|$, depending only on $M$, as defined and studied in \cite[Section 4]{jacobi}, to which we refer for a detailed analysis, while $\textrm{err}^2_{M,N} := \left|  \pi_{f}^{(M)} - \widehat{\pi}_{f}^{(M,N)} \right|$ is the quantization error, on which we focus in the remaining part of this section. 

\begin{remark}
The quantization error $\textrm{err}^2_{M,N}$ formally depends on $M$ and $N$. In our analysis, we consider that the value of $M$ is fixed. In Section \ref{sec:numerical}, before starting numerical investigations, we will choose it so that the truncation error in plain-vanilla option prices can be considered negligible. For this reason, we will simply denote $\textrm{err}^2_{M,N}$ by $\textrm{err}^2_{N}$.
\end{remark}

We study now $\textrm{err}^2_{N}$ via an intermediate lemma and a theorem.

\begin{lemma}\label{lem:error}
The quantization error satisfies
$$
\textrm{err}^2_{N} 
\leq  \left| \left| S_T^{(M)} - \widehat{S}_T^{(M,N)} \right| \right|_{2},
$$%
where 
$S_T^{(M)} := e^{ X_T^{(M)} }$ and $\widehat{S}_T^{(M,N)} $ is the $N$-quantizer relative to ${S}_T^{(M)} $.
\end{lemma}

\begin{proof}
See Appendix \ref{appA}.
\end{proof}

Now we are ready to size the quantization error. As known from Zador Theorem, see Equation \eqref{eq:zador}, the distance $|| S_T^{(M)} - \widehat{S}_T^{(M,N)} ||_{2}$ has an asymptotic linear decay when $N$ goes to infinity. A precise bound for the limit of the error is provided in the following theorem, whose proof is inspired by the one of  \cite[Theorem 2.11]{CFG}. We stress here that the recent error estimates given in \cite[Theorem 2.11]{CFG} were obtained in a different setting and under \textit{ad hoc} assumptions  (the second order moment of $S_T^{(M)}$ was required to be finite and the density of $S_T^{(M)}$ at $0$ and at $+ \infty$ had polynomial behavior), that can be relaxed here, thanks to the explicit form of the density $h_T^{(M)}$ of $S_T^{(M)}$. A thorough study of the error is also present in \cite{FS2019}, in the field of conic finance.



\begin{theorem}\label{thm:quantError} {\bf (Quantization Error Estimate)}\\
In our setting, for any given $M > 0$, we have:
\begin{equation}
\lim_{N \rightarrow  +\infty } \ N \  \textrm{err}^2_{N}  \ \leq \frac{ \left| \left|  h_{T}^{(M)} \right| \right|^{\frac{1}{2}}_{\frac{1}{3}}}{2 \sqrt{3}}
\end{equation}%
where $h_T^{(M)}$ is the density of $S_T^{(M)} = e^{ X_T^{(M)} }$, $h_{T}^{(M)}(s) = \frac{g_{T}^{(M)} (\ln(s))}{s} $ for $s \in (0, + \infty)$. 
\end{theorem}

\begin{proof}
See Appendix \ref{appA}.
\end{proof}

\section[Pricing exotics: multidimensional RMQ]{Pricing exotics: multidimensional RMQ}
\label{sec:secondQ}

The quantization procedure described in the previous section, and based on the polynomial nature of the $(V,X)$, may encounter dimensionality problems when pricing exotic options. In this section we propose an alternative quantization procedure, known as recursive marginal quantization, which in the literature has been discussed for a wide class of stochastic volatility models (see e.g. \cite{CalFioGra2} and \cite{CalFioGraRisk2}), and we extend it to a multidimensional setting.

The innovation behind this approach is twofold. First, the algorithm here is designed for systems of SDEs of any dimension $d$, while the approach used in \cite{CalFioGra2} and \cite{CalFioGraRisk2} is tailored for systems of dimension 2. So, what we present here is a compact and robust alternative to \cite{FiorinSagna}.  The second innovation is in the computational cost required by the execution of the algorithm, which is reduced by a factor $5$ with respect to \cite{CalFioGra2}, see Section \ref{sec:numerical}. Indeed, until now the volatility process was quantized independently, while the price process was then discretized using the information from the approximation of the variance. In the formulation that we propose here, the vector volatility-price process is quantized simultaneously in all its components. This allows, in the end, to more compact formulas and to a more efficient and parsimonious numerical routine.
Here, we consider the quantization of a system of SDEs. We will present a general framework, that we will then apply to the case of the SVJ model. Let us consider the following $d$-dimensional SDE:
\begin{equation}\label{sysSDEs}
d \mathbf{X}_{t} = \mu\left(t, \mathbf{X}_{t} \right) dt + \Sigma\left( t, \mathbf{X}_{t}\right)d\mathbf{W}_{t}, \quad \mathbf{X}_{0}=\mathbf x_{0}.
\end{equation}%
where $\mu: \mathbb{R_+} \times \mathbb{R}^{d} \longrightarrow \mathbb{R}^{d}$, $\Sigma: \mathbb{R_+} \times \mathbb{R}^{d} \longrightarrow  \mathbb{R}^{d}  \times \mathbb{R}^{q}$ and $\mathbf{W}$ is a $q$-dimensional Brownian motion. We suppose that $\mu$ and $\Sigma$ are sufficiently regular so that to ensure existence and uniqueness of a solution to the SDE \eqref{sysSDEs}.

Let us now fix a time discretization grid $t_{k}= k \Delta, k=0, \dots, L$, with $ \Delta= \frac{T}{L}$, where $T$ is a given maturity, $\Delta$ is the time step size and $L$ is the number of discretization points of the time grid. A general discretization scheme can be written in the following iterative form:
\begin{equation}\label{scheme1}
\widetilde{\mathbf{X}}_{t_{k+1}} = A\left(t_{k},\Delta, \widetilde{\mathbf{X}}_{t_{k}}\right) + B\left(t_{k},\Delta, \widetilde{\mathbf{X}}_{t_{k}}, \Delta \mathbf{W}_{t_{k}} \right), \quad \widetilde{\mathbf{X}}_{t_{0}} = x_{0},
\end{equation}%
where $A: \mathbb{R_+} \times \mathbb{R_+} \times \mathbb{R}^{d} \longrightarrow \mathbb{R}^{d} $ and $B: \mathbb{R_+} \times \mathbb{R_+} \times \mathbb{R}^{d} \times \mathbb{R}^{q} \longrightarrow \mathbb{R}^{d}$ depend on the discretization scheme and $\Delta \mathbf{W}_{t_{k}} = \frac{1}{\sqrt{\Delta}} \left( \mathbf{W}_{t_{k+1}}  -  \mathbf{W}_{t_{k}}  \right)$ is a $q$-dimensional Gaussian vector with mean $\mathbf{0}$ and variance-covariance matrix $I_{q}$.

Depending on the time discretization scheme in use, it is possible to know the law of $(\widetilde{\mathbf{X}}_{t_{k+1}} | \widetilde{\mathbf{X}}_{t_{k}}  )$, that clearly depends on $A$ and $B$. In particular, in the case of the Euler-Maruyama (or simply Euler) scheme, that we will choose, $( \widetilde{\mathbf{X}}_{t_{k+1}} | \{ \widetilde{\mathbf{X}}_{t_{k}} = \mathbf x \} ), \mathbf x \in \mathbb R^d,$ has a multivariate Gaussian distribution, while in the case of the Milstein scheme it has a generalized Chi-squared distribution. For higher order schemes, the conditional distribution has to be determined on a case by case basis.

\subsection{Mathematical Foundation of the Algorithm}
\label{Eur}

Henceforth we consider the Euler scheme, so that, conditioning on $\{\widetilde{\mathbf{X}}_{t_{k}} = \mathbf x \}$, we have (recall Equation \eqref{scheme1})
\begin{equation}
A\left(t_{k},\Delta, \mathbf x \right) = \mathbf x + \mu(t_{k}, \mathbf x) \Delta, \quad \quad B\left(t_{k},\Delta,  \mathbf x, \Delta \mathbf{W}_{t_{k}} \right) = \sqrt{\Delta} \Sigma (t_{k}, \mathbf x) \Delta \mathbf{W}_{t_{k}},
\end{equation}%
and the following lemma holds:
\begin{lemma}
For every $0 \le k \le L$, conditionally on the event $\{\widetilde{\mathbf{X}}_{t_{k}} = \mathbf x \}$, the random vector $\widetilde{\mathbf X}_{t_{k+1}}$ is Gaussian:
\begin{equation}
\mathcal L (\widetilde{\mathbf{X}}_{t_{k+1}} | \{ \widetilde{\mathbf{X}}_{t_{k}} =\mathbf x  \})  \sim \mathcal{N} \left( \mathbf x + \mu(t_{k}, \mathbf x) \Delta,  \Delta (\Sigma \Sigma^{T})(t_{k}, \mathbf x) \right).
\end{equation}%
In particular, 
if $\mathbf{X}_{t} = \left( X_{t}^{1}, \dots, X_{t}^{d} \right)$ and $\widetilde{\mathbf{X}}_{t_{k}} = \left( \widetilde{X}_{t_{k}}^{1}, \dots, \widetilde{X}_{t_{k}}^{d} \right)$ for $k=0, \dots, L$, and $\mathbf x=\left(x^{1}, \dots, x^{d}\right)$, we have that for every $i=1, \dots, d$
\begin{equation}\label{gaussSDE}
\mathcal L (\widetilde{X}_{t_{k+1}}^{i}| \{ \widetilde{\mathbf{X}}_{t_{k}} = \mathbf x \} ) \sim \mathcal{N} \left( m_{i}(t_{k}, \mathbf x) , \varsigma_{i}(t_{k}, \mathbf x) \right),
\end{equation}%
where 
$$
m_{i}(t_{k}, \mathbf x):=x^{i} + \mu_{i}(t_{k}, \mathbf x) \Delta 
$$%
is the $i$-th component of the vector $\mathbf x + \mu(t_{k}, \mathbf x)$ and 
$
\varsigma_{i}(t_{k}, \mathbf x) 
$
is the $i$-th diagonal element of the (symmetric) matrix $\Delta \Sigma \Sigma^{T}$.
\end{lemma}

It is then possible to write the distribution of $\widetilde{X}_{t_{k+1}}^{i}$ in a closed form:
\begin{equation}
\label{depgauss}
\mathbb{P}\left( \widetilde{X}_{t_{k+1}}^{i} \in d x_{k+1}^{i} \right) = \int_{\mathbb{R}^{d}} \phi_{m_{i}(t_{k},\mathbf x_{k}) , \varsigma_{i}(t_{k},\mathbf x_{k})}\left(x_{k+1}^{i}\right) \mathbb{P}\left( \widetilde{\mathbf{X}}_{t_{k}} \in d \mathbf x_{k} \right),
\end{equation}%
where $\phi_{m,\varsigma}$ is the probability density function of a one dimensional Gaussian variable with mean $m$ and variance $\varsigma$.

Let us fix a quantization grid $\mathbf{\Gamma}_{i,k+1}=\left\{\gamma_{i,k+1}^{1}, \dots, \gamma_{i,k+1}^{N}\right\}$ of size $N$ relative to $\widetilde{X}_{t_{k+1}}^{i}$. The distortion function associated with $\mathbf{\Gamma}_{i,k+1}$  reads
\begin{equation}
\label{dist}
D_{i,k+1}\left(\mathbf{\Gamma}_{i,k+1}\right) = \sum_{j=1}^{N} \int_{C_{j}\left(\mathbf{\Gamma}_{i,k+1}\right)} \left(x_{k+1}^{i} - \gamma_{i,k+1}^{j} \right)^{2} \mathbb{P}\left( \widetilde{X}_{t_{k+1}}^{i} \in d x_{k+1}^{i} \right)
\end{equation}%
where ${ \left( C_{j}\left(\mathbf{\Gamma}_{i,k+1}\right) \right)}_{j=1, \dots, N}$ is the Voronoi tessellation associated with the grid $\mathbf{\Gamma}_{i,k+1}$.

It is now possible to write the recursive quantization algorithm. Having quantized every $i$-th component of the vector $\widetilde{\mathbf{X}}_{t_{k}}$, via an $N^{i}$-dimensional grid, it is possible to approximate the distribution in \eqref{depgauss} as
\begin{equation}\label{depgaussapp}
\mathbb{P}\left( \widetilde{X}_{t_{k+1}}^{i} \in d x_{k+1}^{i} \right) \approx \sum_{j_{1}=1}^{N^{1}} \cdots  \sum_{j_{d}=1}^{N^{d}}\phi_{m_{i}(t_{k},x_{1,k}^{j_{1}}, \dots, x_{d,k}^{j_{d}}) , \varsigma_{i}(t_{k},x_{1,k}^{j_{1}}, \dots, x_{d,k}^{j_{d}})}\left(x_{k+1}^{i}\right)\mathbb{P}\left( \widetilde{\mathbf{X}}_{t_{k}} = \left( x_{1,k}^{j_{1}}, \dots, x_{d,k}^{j_{d}} \right)\right),
\end{equation}%
where $x_{\ell,k}^{j_{\ell}}$ corresponds to the $j_{\ell}$-th element of the optimal quantization grid of the $\ell$-th component of the vector  $\widetilde{\mathbf{X}}_{t_{k}}$. It is immediate to see that it is possible to compute in closed form also the distribution of the vector  $\widetilde{\mathbf{X}}_{t_{k+1}}$: indeed, we have that
\begin{equation}
\label{depgaussapp2}
\mathbb{P}\left( \widetilde{\mathbf{X}}_{t_{k+1}} \in d x_{k+1} \right) \approx \sum_{j_{1}=1}^{N^{1}} \cdots  \sum_{j_{d}=1}^{N^{d}}\bar{\phi}_{m(t_{k},x_{1,k}^{j_{1}}, \dots, x_{d,k}^{j_{d}}) , \varsigma(t_{k},x_{1,k}^{j_{1}}, \dots, x_{d,k}^{j_{d}})}\left(x_{k+1} \right) \mathbb{P}\left( \widetilde{\mathbf{X}}_{t_{k}} = \left( x_{1,k}^{j_{1}}, \dots, x_{d,k}^{j_{d}} \right)\right),
\end{equation}%
where $\bar{\phi}$ is the density function of a $d$-dimensional Gaussian random variable with mean $m(t_{k},x_{1,k}^{j_{1}}, \dots, x_{d,k}^{j_{d}}) = \left(x_{1,k}^{j_{1}}, \dots, x_{d,k}^{j_{d}}\right) + \mu\left(t_{k},x_{1,k}^{j_{1}}, \dots, x_{d,k}^{j_{d}}\right)$ and variance-covariance matrix $\varsigma(t_{k},x_{1,k}^{j_{1}}, \dots, x_{d,k}^{j_{d}}) = \Delta  (\Sigma \Sigma^{T}) \left(t_{k}, x_{1,k}^{j_{1}}, \dots, x_{d,k}^{j_{d}}\right) $.

Having computed all these elements, it is possible to obtain the (approximate) distortion function \eqref{dist}, its gradient and its Hessian function and to implement the Newton-Raphson algorithm as in \cite{CalFioGra} and \cite{CalFioGra2}.

\subsection{Recursive Quantization of the SVJ Model}
\label{rec:SVJ}

We focus now on the application of the arguments in Section \ref{Eur} to the specific model considered. We consider the Euler scheme of the price $S$, instead of the log price $X$, since quantizing $S$ instead of $X$ is crucial if we want to be in the setting of Section \ref{sec:err} devoted to the study of the numerical error of our procedure. Using the notation of the previous section, 
$\mathbf{X}_{t}=\left(V_{t},S_{t}\right)$ and $\widetilde{X}_{t_{k}}=\left(\widetilde V_{t_{k}},\widetilde S_{t_{k}}\right)$
, and the Euler scheme reads
\begin{equation}\label{rec:Euler}
\begin{pmatrix} \widetilde{V}_{t_{k+1}} \\ \widetilde{S}_{t_{k+1}} \end{pmatrix} = \begin{pmatrix} \widetilde{V}_{t_{k}} \\ \widetilde{S}_{t_{k}} \end{pmatrix} + \begin{pmatrix}  \kappa\left( \theta - \widetilde{V}_{t_{k}} \right) \Delta \\   \left( r - \delta \right) \Delta  \end{pmatrix} + \sqrt{\Delta} \begin{pmatrix}  \sigma \sqrt{Q\left( \widetilde{V}_{t_{k}} \right) } & 0  \\  \rho \widetilde{S}_{t_{k}} \sqrt{ Q\left( \widetilde{V}_{t_{k}} \right) } & \widetilde{S}_{t_{k}} \sqrt{ \widetilde{V}_{t_{k}} - \rho^{2} Q\left( \widetilde{V}_{t_{k}} \right)} \end{pmatrix} \begin{pmatrix} \Delta W^{1}_{k} \\ \Delta W^{2}_{k} \end{pmatrix}.
\end{equation}%

We have then that
\begin{equation}\label{dep:var}
\mathbb{P}\left( \widetilde{V}_{t_{k+1}} \in d v_{k+1} \right) = \int_{\mathbb{R}} \int_{\mathbb{R}} \phi_{m_{1}(t_{k},v_{k},s_{k}) , \varsigma_{1}(t_{k},v_{k},s_{k})}\left(v_{k+1}\right) \mathbb{P}\left( \widetilde{V}_{t_{k}} \in d v_{k}, \widetilde{S}_{t_{k}} \in d s_{k} \right),
\end{equation}%
where $m_{1}(t_{k},v_{k},s_{k}) = v_{k} +\kappa\left( \theta - v_{k} \right) \Delta = m_{1}(t_{k},v_{k}) $ and $\varsigma_{1}(t_{k},v_{k},s_{k}) = \sigma^{2} Q(v_{k})=\varsigma_{1}(t_{k},v_{k})$. 
In the case of the price process
\begin{equation}\label{dep:price}
\mathbb{P}\left( \widetilde{S}_{t_{k+1}} \in d s_{k+1} \right) = \int_{\mathbb{R}} \int_{\mathbb{R}} \phi_{m_{2}(t_{k},v_{k},s_{k}) , \varsigma_{2}(t_{k},v_{k},s_{k})}\left(s_{k+1}\right) \mathbb{P}\left( \widetilde{V}_{t_{k}} \in d v_{k}, \widetilde{S}_{t_{k}} \in d s_{k} \right),
\end{equation}%
where $m_{2}(t_{k},v_{k},s_{k}) = s_{k} + \left( r - \delta \right) \Delta$ and $\varsigma_{2}(t_{k},v_{k},s_{k}) = \left(s_{k}\right)^{2} v_{k}$. Moreover, we notice that, since $m_{1}$ and $\varsigma_{1}$ do not depend on $s_{k}$, we can simplify \eqref{dep:var}:
\begin{equation}\label{dep:var2}
\mathbb{P}\left( \widetilde{V}_{t_{k+1}} \in d v_{k+1} \right) = \int_{\mathbb{R}} \phi_{m_{1}(t_{k},v_{k}) , \varsigma_{1}(t_{k},v_{k})}\left(v_{k+1}\right) \mathbb{P}\left( \widetilde{V}_{t_{k}} \in d v_{k} \right). 
\end{equation}%



This allows to use the technique developed in \cite{PagSagna} and \cite{CalFioGra} for the quantization of the variance process, which is one dimensional and it can be discretized independently of $S$. On the other hand, of course, the quantization grids for $S$ will depend on the ones for $V$.

We now give an idea on how it is possible to recursively obtain the quantization grids $\mathbf{\Gamma}_{1,k} =: \mathbf{\Gamma}_{V,k}$ and $\mathbf{\Gamma}_{2,k} =: \mathbf{\Gamma}_{S,k}, k=1,\dots,L$. 
We suppose that the cardinality of the grids is fixed: $| \mathbf{\Gamma}_{1,k} | = N^V$ and $ | \mathbf{\Gamma}_{2,k} | = N^S$, for every $k $. Moreover, we recall that the quantization grids at time $t_0=0$, $\mathbf{\Gamma}_{V,0}$ and $\mathbf{\Gamma}_{S,0}$, are vectors whose components correspond, respectively, with $v_0$ and $S_0$. 

Let us assume now that we have computed the optimal grids for the variance and the price process, namely $\mathbf{\Gamma}_{V,k} =\left\{v_k^{1}, \dots, v_k^{N_{V}} \right\}$  for the variance process and   $\mathbf{\Gamma}_{S,k} =\left\{s_k^{1}, \dots, s_k^{N_{S}} \right\}$ for the price process, up to time $t_{k}$ and that we want to obtain  $\mathbf{\Gamma}_{V,k+1}$ and $\mathbf{\Gamma}_{S,k+1}$. To do this, we look for the zeros of the gradient of the distortion function \eqref{dist} when the probability \eqref{depgaussapp2} takes the form
\begin{equation}\label{jointPJacobi}
\mathbb{P}\left( \widetilde{V}_{t_{k+1}} \in d v_{k+1}, \widetilde{S}_{t_{k+1}} \in d s_{k+1} \right) \approx \sum_{i=1}^{N^{V}} \sum_{j=1}^{N^{S}} \bar{\phi}_{m(t_{k},v_k^{i},s_{k}^{j}),\varsigma(t_{k},v_k^{i},s_{k}^{j})}(v_{k+1},s_{k+1})  \mathbb{P}\left( \widetilde{V}_{t_{k}} = v_{k}^{i}, \widetilde{S}_{t_{k}} = s_{k}^{j} \right),
\end{equation}%
where $\bar{\phi}$ is the density of a bivariate Gaussian with mean 

$$
m(t_{k},v_k^{i},s_{k}^{j}) = \begin{pmatrix} v_{k}^{i} +\kappa\left( \theta - v_{k}^{i} \right) \Delta  \\    s_{k}^{j} + \left( r - \delta \right) \Delta\end{pmatrix}
$$%
and variance 
$$
\varsigma(t_{k},v_k^{i},s_{k}^{j}) = \Delta \begin{pmatrix}   \sigma^{2} Q(v_{k}^{i}) & \rho \sigma s_{k}^{j}  Q(v_{k}^{i})  \\   \rho \sigma s_{k}^{j} Q(v_{k}^{i})  &   \left(s_{k}^{j}\right)^{2} v_{k}^{i}  \end{pmatrix}.
$$



\begin{remark} {\bf (Calculation of Transition Probabilities)}\\ \label{rem:ptrans}
As a byproduct of recursive quantization, we instantaneously get for free also the transition probabilities. Indeed, from \eqref{jointPJacobi} we immediately have the transition densities
$$
\mathbb{P}\left( \widetilde{V}_{t_{k+1}} \in d v_{k+1}, \widetilde{S}_{t_{k+1}} \in d s_{k+1} | \widetilde{V}_{t_{k}} = v_{k}^{i}, \widetilde{S}_{t_{k}} = s_{k}^{j}  \right) \approx \bar{\phi}_{m(t_{k},v_k^{i},s_{k}^{j}),\varsigma(t_{k},v_k^{i},s_{k}^{j})}(v_{k+1},s_{k+1})  ,
$$%
for $i=1,\dots,N^{V}$ and $j=1,\dots,N^{S}$.
\end{remark}

\section{Numerical Results}
\label{sec:numerical}

In this section we present numerical results on pricing of European and Bermudan options. Polynomial quantization is only used to price vanilla options
, while recursive marginal quantization, allowing for an immediate approximation of the transition probabilities, is exploited to price both European and Bermudan derivatives. 
It is noteworthy that the pricing of vanilla and exotic options still represents a challenge from a numerical point of view when the dimension of the process considered is strictly greater than one. 
Numerical results have been obtained in Matlab 9.2, with a CPU 2.7 GHz and 4 Gb memory computer.

\begin{remark}{\bf (Barrier and Asian options)}\\
If interested in different types of path dependent options, such as barrier or Asian, we refer the reader to the methodology developed in  \cite{BCLP15}, where, on top of a grid for the underlying process at every intermediate date (together with transition probabilities from one time step to another) a backward Monte Carlo procedure is applied to price barrier, Asian and auto-callable options.
\end{remark}

We choose the following values for the mean and the standard deviation of the Gaussian weight function $w$ (also recall Remark \ref{rem:ell_in_L^2}), which are the same as in \cite{jacobi}: 
$$
\sigma_{w}=\sqrt{\frac{v_{max}T}{2}} + 10^{-4}, \quad \mu_{w}=\mathbb{E}\left[ X_{T} \right] 
$$%
and in all the numerical examples we will consider the parameters in Table \ref{param}:
\begin{table}[htp]
	\begin{center}
		\begin{tabular}{|c|c|c|c|}
			\hline
			$\kappa=1.7$ & $\theta=0.06$ & $\sigma=0.5$ & $\rho=-0.5$ \\
			\hline
			$V_{0}=0.1 $ &$v_{min}=10^{-2}$ & $v_{max}=1$ & $r=0.04$ \\
			\hline
			$\delta=0 $ &$S_{0}=100$ &$T=1$ & $M=100$ \\
			\hline
		\end{tabular}
	\end{center}
	\caption{Parameters of the SVJ model.}\label{param}
\end{table}

\subsection{The choice of $M$}

Before showing our results, we discuss the choice of $M$. This is crucial, since it is known that the density $g_T^{(M)}$ introduced in Equation \eqref{g_T^(M)} it is not guaranteed to be non-negative (see e.g. \cite[Figure 3]{jacobi}).  In the literature this is a well-known issue and it is due to the polynomial nature of the terms involved in the Gram Charlier expansion of the density. While conditions on the expansion coefficients which ensure positivity of the density have been found in the case when the series is truncated at the fourth order moment (see e.g. \cite{JR99}), no results are known, to our knowledge, in the general case. Various tests and alternative approaches and expansions have been proposed and we cite, among them, \cite{RE2007}, \cite{LMS2009}, \cite{NP2012}, \cite{CD2017} and \cite{Sc2013}.

Investigating this problem in details is out of the scope of the paper, since changing the series expansion would lead to a brand new contribution, so in this subsection we motivate our choice of $M$, 
based on an analysis of the produced vanilla prices. We proceed by computing the prices of call options for different strikes via a Monte Carlo simulation with $2 \cdot 10^6$ paths and antithetic variates. In Table \ref{mc_lucio} we show numerical results along with a $95 \%$ confidence interval for Monte Carlo prices.

\begin{table}[htp]
	\begin{center}
		\begin{tabular}{|c|c|c|c|}
			\hline
			Strike & CI lower bound & MC price & CI upper bound \\
			\hline			
			$80$& $  25.8761  $ &$  25.8984   $ &$ 25.9207 $\\
			$85$&  $ 22.1012   $ &$ 22.1224   $ &$ 22.1436 $\\
			$90$&  $ 18.5964   $ &$ 18.6164  $ &$  18.6364$\\
			$95$&  $ 15.3991    $ &$15.4178   $ &$ 15.4364$\\
			$100$&  $ 12.5398    $ &$12.5570   $ &$ 12.5742$\\
			$105$&  $ 10.0396   $ &$ 10.0552   $ &$ 10.0709$\\
			$110$&  $  7.9047   $ &$  7.9188    $ &$ 7.9329$\\
			$115$&   $ 6.1252    $ &$ 6.1379   $ &$  6.1505$\\
			$120$&   $ 4.6775    $ &$ 4.6886    $ &$ 4.6998$\\
			\hline
		\end{tabular}
	\end{center}
	\caption{Monte Carlo prices and $95\%$ confidence interval of Call options obtained with $2 \cdot 10^6$ simulations and antithetic variates.}\label{mc_lucio}
\end{table}

In order to choose $M$, we then evaluate Call options via the procedure introduced in \cite{jacobi}, for the same strike values in Table \ref{mc_lucio} and for different values of $M$ and we check for which values of $M$ the obtained price belongs to the $95 \%$ Monte Carlo confidence interval in Table \ref{mc_lucio} for every strike. Results are presented in Table \ref{pol_lucio} and they show that a safe choice is $M \ge 90$. In the following, we choose the more conservative value of $M=100$. Notice that this choice is in line with \cite{jacobi}.

\begin{table}[htp]
	\begin{center}
		\begin{tabular}{|c|c|c|c|c|c|c|c|c|c|}
			\hline
			Strike & $M=20$&  $M=30$&  $M=40$&  $M=50$&  $M=60$&  $M=70$&  $M=80$&  $M=90$&  $M=100$  \\	
			\hline
			$80$& $    25.7122  $   &$  25.7899$   &$    25.8457$   &$    25.8759$   &$    25.8906$   &$    25.8970$   &$    25.8992$   &$    25.8995$   &$    25.8991 $\\
			$85$& $  22.0032   $   &$ 22.0213$   &$    22.0586$   &$   22.0857$   &$    22.1023$   &$    22.1119$   &$    22.1171$   &$    22.1198$   &$    22.1211$\\
			$90$& $   18.6058  $   &$  18.5559  $   &$  18.5642  $   &$  18.5794  $   &$  18.5919  $   &$  18.6006 $   &$   18.6064  $   &$  18.6101 $   &$   18.6125$\\
			$95$& $  15.5373  $   &$  15.4222  $   &$  15.3976  $   &$  15.3958   $   &$ 15.3996 $   &$   15.4040  $   &$  15.4078 $   &$   15.4107  $   &$  15.4129$\\
			$100$& $  12.8039  $   &$  12.6355 $   &$   12.5806  $   &$  12.5612  $   &$  12.5546  $   &$  12.5528  $   &$  12.5529  $   &$  12.5536  $   &$  12.5544$\\
			$105$& $  10.4018 $   &$   10.1979 $   &$   10.1207 $   &$   10.0870$   &$    10.0712  $   &$  10.0634  $   &$  10.0595  $   &$  10.0576  $   &$  10.0566$\\
			$110$& $   8.3189 $   &$    8.1007  $   &$   8.0121  $   &$   7.9703 $   &$    7.9487  $   &$   7.9369  $   &$   7.9301 $   &$    7.9261  $   &$   7.9237$\\
			$115$& $   6.5370  $   &$   6.3259 $   &$    6.2381  $   &$   6.1954  $   &$   6.1726 $   &$    6.1597 $   &$    6.1520 $   &$    6.1473 $   &$    6.1442$\\
			$120$& $   5.0336  $   &$   4.8491  $   &$   4.7734  $   &$   4.7370  $   &$   4.7176  $   &$   4.7066  $   &$   4.7001   $   &$  4.6961  $   &$   4.6934$\\
			\hline
		\end{tabular}
	\end{center}
	\caption{Pricing of Call options via the methodology in \cite{jacobi}, for different strikes and different values of $M$.}\label{pol_lucio}
\end{table}

\subsection{Polynomial Quantization}
We use the technique developed in Section \ref{sec:poly}, and we compute the quantization grid associated to the log price process at time $T$, i.e. we approximate $X_{T}$ using an optimal grid $\Gamma^{*}=\{ x_{1}^*, \dots, x_{N}^* \}$. The price of a Call option with maturity $T$ and strike $K$ is then approximated as
$$
e^{- r T}\mathbb{E}\left[  \left(e^{X_{T}}-K\right)^{+} \right] \approx e^{- r T}\sum_{i=1}^{N} \left(e^{x_{i}^*}-K\right)^{+} \mathbb{P} \left(  X_{T} \in C_{i}(\Gamma^{*})  \right),
$$%
where $C_{i}(\Gamma^{*}) = \left[ \frac{x_{i-1}^* + x_{i}^*}{2}, \frac{x_{i}^* + x_{i+1}^*}{2} \right]$ and the weights are given by \eqref{weights}. The results in Table \ref{vanilla:pol} show that the quantization technique is accurate, when compared with the benchmark price, given by \cite{jacobi}. Moreover, the computational cost is comparable to the execution time declared in \cite{jacobi} ((see Figure 8)): computing Hermite moments requires the same time declared therein, getting the quantization grids costs $0.94$ seconds and the price computation is immediate, so that the total computational times are equivalent. 

\begin{table}[htp]
\begin{center}
\begin{tabular}{|c|c|c|c|}
\hline
\multicolumn{1}{|c|}{Strike} & \multicolumn{1}{|c|}{Benchmark price} & \multicolumn{1}{|c|}{Quantization price} & \multicolumn{1}{|c|}{Relative error (\%)} \\
\hline
\multicolumn{1}{|l|}{$K=    80 $} & $25.8991 $ &$ 25.8646 $&$ 0.1332 $\\
\multicolumn{1}{|l|}{$K=    85 $} & $22.1211 $ &$ 22.0980 $&$ 0.1044 $\\
\multicolumn{1}{|l|}{$K=    90 $} & $18.6125 $ &$ 18.5866 $&$ 0.1391 $\\
\multicolumn{1}{|l|}{$K=    95 $} & $15.4129 $ &$ 15.3648 $&$ 0.3121 $\\
\multicolumn{1}{|l|}{$K=   100 $} & $12.5544 $ &$ 12.4710 $&$ 0.6643 $\\
\multicolumn{1}{|l|}{$K=   105 $} & $10.0566 $ &$ 10.0165 $&$ 0.3987 $\\
\multicolumn{1}{|l|}{$K=   110 $} & $7.9237 $ &$ 7.9066 $&$ 0.2158$\\
\multicolumn{1}{|l|}{$K=   115 $} & $6.1442 $ &$ 6.1093 $&$ 0.5680$\\
\multicolumn{1}{|l|}{$K=   120 $} & $4.6934 $ &$ 4.6626 $&$ 0.6562 $\\
\hline
\end{tabular}
\end{center}
\caption{Pricing comparison between the benchmark price and the price obtained via polynomial quantization of a European Call option for the SVJ model with parameters  as in Table \ref{param}. The quantization grids have size $N=20$.}\label{vanilla:pol}
\end{table}

\subsection{Recursive Quantization}\label{sec:RecQuant}

We use the methodology implemented in Section \ref{rec:SVJ}. Note that we do not exploit the fact that $S$ is the exponential of a polynomial process, but we construct the optimal quantizers starting from the Euler scheme  \eqref{rec:Euler}. We then compute, at every time step $t_{k}$, for $k=1, \dots, L$, such that $t_{L}=T$, the quantization of the price process $S$ at time $t_{k}$, that we call $\widehat{S}_{t_{k}}$. 
In order to price a European call option with strike $K$ and maturity $T$ we need only $\Gamma_{S,L}^{*} = \{s^{L}_{1}, \dots, s_{N}^{L} \}$,  the optimal quantization grid associated to $\widehat{S}_{T}$, and we have the following approximation:
$$
e^{- r T}\mathbb{E}\left[  \left(e^{X_{T}}-K\right)^{+} \right] \approx \sum_{j=1}^{N^{S}} \left( s^{j}_{L} - K \right)^{+} \mathbb{P} \left( \widehat{S}_{T} = s^{j}_{L} \right),
$$%
where the weights are computed using \eqref{dep:price}. The results in Table \ref{vanilla:rec} show that recursive quantization is efficient when compared to the benchmark methodology by \cite{jacobi} and that it is a good alternative to polynomial quantization (see Table \ref{vanilla:pol}). Moreover, recursive quantization does not require the computation of Hermite moments, so the computational cost here is only relative to the computation of the quantization grids (recall, nevertheless, that here we have to both discretize the volatility and the price process), which corresponds to $2.24$ seconds, with $N^S=20, N^V=10$ and $L=12$ (so, a total of $360$ points).  

\begin{table}[htp]
\begin{center}
\begin{tabular}{|c|c|c|c|}
\hline
\multicolumn{1}{|c|}{Strike} & \multicolumn{1}{|c|}{Benchmark price} & \multicolumn{1}{|c|}{Quantization price} & \multicolumn{1}{|c|}{Relative error (\%)} \\
\hline
\multicolumn{1}{|l|}{$K=    80 $} & $25.8991 $ &$ 25.9082 $&$ 0.0351$\\
\multicolumn{1}{|l|}{$K=    85 $} & $22.1211$ &$ 22.1462 $&$ 0.1135 $\\
\multicolumn{1}{|l|}{$K=    90 $} & $18.6125 $ &$ 18.6430 $&$ 0.1639 $\\
\multicolumn{1}{|l|}{$K=    95 $} & $15.4129 $ &$ 15.4395 $&$ 0.1726 $\\
\multicolumn{1}{|l|}{$K=   100 $} & $12.5544 $ &$ 12.5677 $&$ 0.1059 $\\
\multicolumn{1}{|l|}{$K=   105 $} & $10.0566 $ &$ 10.0789 $&$ 0.2217$\\
\multicolumn{1}{|l|}{$K=   110 $} & $7.9237 $ &$ 7.9508 $&$ 0.3420 $\\
\multicolumn{1}{|l|}{$K=   115 $} & $6.1442 $ &$ 6.1692 $&$ 0.4069 $\\
\multicolumn{1}{|l|}{$K=   120 $} & $4.6934 $ &$ 4.7106 $&$ 0.3665 $\\
\hline
\end{tabular}
\end{center}
\caption{Pricing comparison between the benchmark price and the price obtained via recursive quantization of a European Call option for the SVJ model with parameters  as in Table \ref{param}. The quantization grids have size ${N^{S}=20}$, ${N^{V}=10}$ for every time step, and $L=12$. 
}\label{vanilla:rec}
\end{table}

\subsection{Bermudan Options}

The advantage of the Recursive Marginal quantization algorithm developed in Section \ref{Eur} is the possibility to price path dependent options, since we approximate the process at every time step of the Euler scheme, and the transition densities are given directly by the algorithm, as shown in \eqref{jointPJacobi}. This motivates us to show an application of this methodology to the pricing of Bermudan options. Pricing such options can be done via a backward procedure on the multinomial tree obtained via quantization, as presented e.g. in \cite[Proposition 2.1]{BP_american_quant}. As a first benchmark used for comparison we consider the Longstaff Schwarz algorithm, as done also in \cite{Filipovic_cubature} when dealing with American Put options in a Jacobi exchange rate model. 
The results in Table \ref{bermudan:rec} show the accuracy of our methodology. The computational cost behind the pricing of Bermudan Options using recursive quantization is the one required by the computation of the quantization grids (which do not depend on the derivative's strike), that is the same declared in Section \ref{sec:RecQuant} ($2.24$ seconds), and the Bermudan backward algorithm, that, due to the size of the grids in our example, is instantaneous. 

\begin{table}[htp]
	\begin{center}
		\begin{tabular}{|c|c|c|c|}
			\hline
			\multicolumn{1}{|c|}{Strike} & \multicolumn{1}{|c|}{Benchmark price} & \multicolumn{1}{|c|}{Quantization price} & \multicolumn{1}{|c|}{Relative error (\%)} \\
			\hline
			\multicolumn{1}{|l|}{$K=    80 $} & $3.0410 $ &$ 2.9984 $&$ 1.3997 $\\
			\multicolumn{1}{|l|}{$K=    85 $} & $4.1040 $ &$ 4.1077 $&$ 0.0899 $\\
			\multicolumn{1}{|l|}{$K=    90 $} & $5.4579 $ &$ 5.5012 $&$ 0.7931 $\\
			\multicolumn{1}{|l|}{$K=    95 $} & $7.1493 $ &$ 7.2222 $&$ 1.0199 $\\
			\multicolumn{1}{|l|}{$K=   100 $} & $9.2192 $ &$ 9.3151 $&$ 1.0404 $\\
			\multicolumn{1}{|l|}{$K=   105 $} & $11.6984 $ &$ 11.8285 $&$ 1.1120 $\\
			\multicolumn{1}{|l|}{$K=   110 $} & $14.6035 $ &$ 14.7564 $&$ 1.0470 $\\
			\multicolumn{1}{|l|}{$K=   115 $} & $17.9352 $ &$ 18.0969 $&$ 0.9015 $\\
			\multicolumn{1}{|l|}{$K=   120 $} & $21.6788 $ &$ 21.8295 $&$ 0.6951 $\\
			\hline
		\end{tabular}
	\end{center}
	\caption{Benchmark price (Longstaff Schwartz) and the price obtained via recursive quantization of a Bermudan Put option for the SVJ model with parameters as in Table \ref{param}. The quantization grids have size ${N^{S}=20}$, ${N^{V}=10}$ for every time step, and $L=12$. 
	}\label{bermudan:rec}
\end{table}

For completeness we also compare our Bermudan option prices with those in \cite[Section 5.2.1]{CKN2018}, the benchmark there being still Longstaff Schwartz. In Table \ref{paramCui} we list the parameters used therein, which we also consider here for comparison. The computational time declared by our competitors for seven strikes is $4.7$ seconds with Matlab 8.5 on a personal computer with an i7-6700 CPU @3.40 GHz. This is analogous to the time required by RMQ to obtain all the prices below.
\begin{table}[htp]
	\begin{center}
		\begin{tabular}{|c|c|c|c|}
			\hline
			$\kappa=3$ & $\theta=0.13$ & $\sigma=0.4$ & $\rho=-0.2$ \\
			\hline
			$V_{0}=0.13 $ &$v_{min}=10^{-2}$ & $v_{max}=0.25$ & $r=0.02$ \\
			\hline
			$\delta=0 $ &$S_{0}=10$ &$T=0.5$ & \\
			\hline
		\end{tabular}
	\end{center}
	\caption{Parameters of the SVJ model in \cite{CKN2018}.}\label{paramCui}
\end{table}

In Table \ref{bermudan:Cui} we show, for seven different strikes: the Longstaff-Schwartz benchmark price, our RMQ price, the price proposed by \cite{CKN2018} obtained via a continuous-time Markov chain approximation and the two absolute errors.

\begin{table}[htp]
	\begin{center}
		\begin{tabular}{|c|c|c|c|c|c|}
			\hline
			\multicolumn{1}{|c|}{Strike} & \multicolumn{1}{|c|}{Benchmark (L.-S.)} & \multicolumn{1}{|c|}{RMQ price} & \multicolumn{1}{|c|}{Cui et al. price} & \multicolumn{1}{|c|}{Abs err RMQ} & \multicolumn{1}{|c|}{Abs err Cui et al.} \\
			\hline
			\multicolumn{1}{|l|}{$K=    8.5 $} & $0.3527 $ & $0.3603$ &$0.3522$ & $0.0076$ & $0.0005$ \\
			\multicolumn{1}{|l|}{$K=    9 $} & $0.5125 $ & $0.5209$ &$0.5139$ & $0.0084$ & $0.0014$ \\
			\multicolumn{1}{|l|}{$K=    9.5 $} & $0.7157 $ & $0.7213$ &$0.7163$ & $0.0056$ & $0.0006$ \\
			\multicolumn{1}{|l|}{$K=    10 $} & $0.9591 $ & $0.9624$ &$0.9599$ & $0.0033$ & $0.0008$ \\
			\multicolumn{1}{|l|}{$K=   10.5 $} & $1.2407 $ & $1.2430$ &$1.2435$ & $0.0023$ & $0.0028$ \\
			\multicolumn{1}{|l|}{$K=   11 $} & $1.5625$ & $1.5620$ &$1.5643$ & $0.0005$ & $0.0018$ \\
			\multicolumn{1}{|l|}{$K=   11.5 $} & $1.9163$ & $1.9156$ &$1.9189$ & $0.0007$ & $0.0026$ \\
			\hline
		\end{tabular}
	\end{center}
	\caption{Benchmark price (Longstaff-Schwartz) provided in \cite{CKN2018}, the price obtained via RMQ quantization, the price in \cite{CKN2018} and the absolute errors for a Bermudan Put option for the SVJ model with parameters as in Table \ref{paramCui}. The quantization grids have size ${N^{S}=40}$, ${N^{V}=10}$ for every time step, and $L=12$. 
	}\label{bermudan:Cui}
\end{table}

The results in Table \ref{bermudan:Cui} confirm (recall Table \ref{bermudan:rec}) the effectiveness of RMQ when pricing path-dependent options in the SJV model.

\section{Conclusion}

In this paper we presented how to efficiently apply quantization techniques to polynomial processes. In particular, we focused on the SVJ model, but our results can be extended to any polynomial model. Our analysis on SVJ quantization provided numerical tools to develop fast exotic option pricing algorithms. We presented two approaches. Firstly, we exploited the polynomial property, and we provided new theoretical results to study the approximation errors. As a result we obtained \textit{ad-hoc} pricing tools for polynomial models 
and we provided numerical examples to assess their goodness with respect to the existing literature. Secondly, we applied RMQ to polynomial processes, by viewing them as a particular class of stochastic volatility processes. This allowed us to price exotic options and numerical examples for Bermudan options were given. Our conclusion is that quantization is a powerful discretization procedure, with respect to precision and speed, whose paradigm can be easily used in the field of polynomial processes. Moreover, when looking at the family of quantization procedures with pricing in view, recursive marginal quantization is undoubtedly the most powerful, given its flexibility and effectiveness.

\appendix

\section{Proofs of the Main Results}
\label{appA}

We here provide the proofs of the main results obtained in the paper.

\subsection[Proof of Theorem 4.1]{Proof of Theorem \ref{thm:masterEq}}

\begin{proof}
First of all notice that, for every $i=1, \dots, N$, the expectation in \eqref{masterSys} can be rewritten as $\mathbb E [f^i(X_T)]$, with $f^{i}(y) : = \left( y - x_{i} \right) 1\!\!1_{y \in \left[ \frac{x_{i-1} + x_{i}}{2}, \frac{x_{i} + x_{i+1}}{2} \right]}$. In order to exploit the polynomial nature of our setting and to use the result in Equation \eqref{exppol}, we need $f^i$ to be in $ L_{w}^{2}$.  We have
\begin{align*}
||f^i ||_{w}^{2} = \int_{\mathbb{R}} \left( f^{i}(y) \right)^{2} w(y) dy & \leq \int_{\mathbb{R}} \left( y - x_{i} \right)^{2} w(y) dy \\
& = \int_{\mathbb{R}}  y^{2} \ w(y) \ dy - 2 x_{i} \int_{\mathbb{R}}  y \  w(y) \ dy + x_{i}^{2} \int_{\mathbb{R}}  w(y) dy \\
& = \sigma_{w}^{2} + \mu_{w}^{2} - 2 x_{i} \mu_{w} + x_{i}^{2}
\end{align*}
which is finite for every $ i=1, \dots, N$. 

We want to compute the expected value in \eqref{masterSys_i}. Using the polynomial property in \eqref{exppol}, we now rewrite it in the form of \eqref{masterpol}, where (recalling Equation \eqref{fourier})
\begin{align*}
f^{i}_{n} & = \int_{\mathbb{R}}  \left( y - x_{i} \right) 1\!\!1_{y \in \left[ \frac{x_{i-1} + x_{i}}{2}, \frac{x_{i} + x_{i+1}}{2} \right]} H_{n}(y) w(y) dy \\
& = \underbrace{\int_{\mathbb{R}} y  1\!\!1_{\left[ \frac{x_{i-1} + x_{i}}{2}, \frac{x_{i} + x_{i+1}}{2} \right]}(y) H_{n}(y) w(y) dy}_{a^{i}_{n}} - x_{i} \underbrace{\int_{\mathbb{R}}1\!\!1_{\left[ \frac{x_{i-1} + x_{i}}{2}, \frac{x_{i} + x_{i+1}}{2} \right]}(y) H_{n}(y) w(y) dy}_{b^{i}_{n}}
\end{align*}%
We focus first on the computation of $a^{i}_{n}$. Let us define
$$
h_{n}(K) = \int_{\mathbb{R}} y  1\!\!1_{\left[ K, \infty \right]}(y) H_{n}(y) w(y) dy ,\\
$$%
then $a_{n}^{i} = h_{n}\left(  \frac{x_{i-1} + x_{i}}{2} \right) - h_{n}\left(  \frac{x_{i} + x_{i+1}}{2} \right) $. When $n=0$ we have, integrating by parts, that
\begin{align*}
h_{0}(K) & = \int_{\mathbb{R}} y  1\!\!1_{\left[ K, \infty \right]}(y) w(y) dy\\
& = \sigma_{w} \phi\left( \frac{K - \mu_{w}}{\sigma_{w}} \right) + \mu_{w} \Phi \left( \frac{\mu_{w}-K}{\sigma_{w}}\right).
\end{align*}%
When $n \geq 1$ we have that
\begin{align*}
h_{n}(K) & = \int_{K}^{\infty} y  H_{n}(y) w(y) dy \\
& = \frac{1}{\sqrt{n !}} \int_{K}^{\infty} y  \mathcal{H}_{n}\left( \frac{y - \mu_{w}}{\sigma_{w}}  \right) \frac{1}{\sigma_{w}}  \phi\left( \frac{y - \mu_{w}}{\sigma_{w}}  \right) dy\\
& = \frac{1}{\sqrt{n !}} \int_{\frac{K - \mu_{w}}{\sigma_{w}}}^{\infty} \left(\sigma_{w}z + \mu_{w}\right)  \mathcal{H}_{n}\left( z  \right)   \phi\left( z \right) dz \\
& = \frac{\sigma_{w}}{\sqrt{n !}} \int_{\frac{K - \mu_{w}}{\sigma_{w}}}^{\infty}z  \mathcal{H}_{n}\left( z  \right)   \phi\left( z \right) dz + \frac{\mu_{w}}{\sqrt{n !}} \int_{\frac{K - \mu_{w}}{\sigma_{w}}}^{\infty}\mathcal{H}_{n}\left( z  \right)   \phi\left( z \right) dz.
\end{align*}%
We exploit the recursive relation between the Hermite polynomials  for $n \geq 1$, : $z \mathcal{H}_{n}(z) = \mathcal{H}_{n+1}(z) + n \mathcal{H}_{n-1}(z)$, in order to get
\begin{align*}
h_{n}(K) & = \frac{\sigma_{w}}{\sqrt{n !}} \int_{\frac{K - \mu_{w}}{\sigma_{w}}}^{\infty} \mathcal{H}_{n+1}\left( z  \right)   \phi\left( z \right) dz + \frac{n \sigma_{w}}{\sqrt{n !}} \int_{\frac{K - \mu_{w}}{\sigma_{w}}}^{\infty} \mathcal{H}_{n-1}\left( z  \right)   \phi\left( z \right) dz + \frac{\mu_{w}}{\sqrt{n !}} \int_{\frac{K - \mu_{w}}{\sigma_{w}}}^{\infty}\mathcal{H}_{n}\left( z  \right)   \phi\left( z \right) dz.
\end{align*}%
The case $n=1$ can be obtained directly using integration by parts, while \cite[Theorem 3.7]{jacobi} proves that, for $n \geq 1$,
$$
\int_{x}^{\infty}\mathcal{H}_{n}\left( z  \right)   \phi\left( z \right) dz = \mathcal{H}_{n-1}\left( x  \right)   \phi\left( x \right),
$$%
so we have the result for $h_{n}(K)$. The $b_{n}^{i}$ coefficients can be computed similarly. In fact, if we define
$$
l_{n}(K) = \int_{\mathbb{R}} 1\!\!1_{\left[ K, \infty \right]}(y) H_{n}(y) w(y) dy,
$$%
then $b_{n}^{i} = l_{n}\left(  \frac{x_{i-1} + x_{i}}{2} \right) - l_{n}\left(  \frac{x_{i} + x_{i+1}}{2} \right) $. The case when $n=0$ is trivial, instead when $n \geq 1$ we have that
\begin{align*}
l_{n}(K) & = \int_{K}^{\infty} H_{n}(y) w(y) dy \\
& = \frac{1}{\sqrt{n !}} \int_{K}^{\infty} \mathcal{H}_{n}\left( \frac{y - \mu_{w}}{\sigma_{w}}  \right) \frac{1}{\sigma_{w}}  \phi\left( \frac{y - \mu_{w}}{\sigma_{w}}  \right) dy\\
& = \frac{1}{\sqrt{n !}} \int_{\frac{K - \mu_{w}}{\sigma_{w}}}^{\infty} \mathcal{H}_{n}\left( z  \right)   \phi\left( z \right) dz \\
& = \frac{1}{\sqrt{n !}} \mathcal{H}_{n-1}\left( \frac{K - \mu_{w}}{\sigma_{w}} \right)   \phi\left(  \frac{K - \mu_{w}}{\sigma_{w}}\right).
\end{align*}%
\end{proof}

\subsection[Proof of Proposition 4.2]{Proof of Proposition \ref{propJacobian}}

\begin{proof}
Remember that $g_{T}$ is the density of $X_{T}$. We can then rewrite $E_{i}(x_{1}, \dots, x_{N})$ as
$$
E_{i}(x_{1}, \dots, x_{N}) = \int_{\frac{x_{i-1}+x_{i}}{2}}^{{\frac{x_{i}+x_{i+1}}{2}}} y g_{T}(y) dy - x_{i} \int_{\frac{x_{i-1}+x_{i}}{2}}^{{\frac{x_{i}+x_{i+1}}{2}}} g_{T}(y) dy .
$$%
This shows that $E_{i}$ depends only on $x_{i-1}$, $x_{i}$ and $x_{i+1}$, so that the Jacobian matrix $J$ is tridiagonal. Moreover the lower diagonal has components:
\begin{align*}
J_{i,i-1}= \frac{ \partial E_{i}}{\partial x_{i-1}}(x_{i-1}, x_{i}, x_{i+1})  & =  - \frac{1}{2}\frac{x_{i-1}+x_{i}}{2} g_{T}\left(\frac{x_{i-1}+x_{i}}{2}\right) + \frac{1}{2} x_{i}  g_{T}\left(\frac{x_{i-1}+x_{i}}{2}\right)   \\
& = \frac{1}{2} \left( \frac{x_{i} - x_{i-1}}{2} \right) g_{T} \left( \frac{x_{i-1} + x_{i}}{2} \right)
\end{align*}%
and the upper diagonal reads:
\begin{align*}
J_{i,i+1}= \frac{ \partial E_{i}}{\partial x_{i+1}}(x_{i-1}, x_{i}, x_{i+1})  & =   \frac{1}{2}\frac{x_{i}+x_{i+1}}{2} g_{T}\left(\frac{x_{i}+x_{i+1}}{2}\right) - \frac{1}{2} x_{i}   g_{T}\left(\frac{x_{i}+x_{i+1}}{2}\right)  \\
& = \frac{1}{2} \left( \frac{x_{i+1} - x_{i}}{2} \right) g_{T} \left( \frac{x_{i} + x_{i+1}}{2} \right).
\end{align*}%
We can deduce immediately that $J_{i,i-1} = J_{i-1,i}$, so that $J$ is also symmetric. Finally the diagonal has components:
\begin{align*}
J_{i,i}= \frac{ \partial E_{i}}{\partial x_{i}}(x_{i-1}, x_{i}, x_{i+1})  & =   \frac{1}{2} \frac{x_{i}+x_{i+1}}{2} g_{T}\left(\frac{x_{i}+x_{i+1}}{2}\right)  -  \frac{1}{2} \frac{x_{i}+x_{i+1}}{2} g_{T}\left(\frac{x_{i}+x_{i+1}}{2}\right)     \\
& - \frac{1}{2} x_{i} \left(     g_{T}\left(\frac{x_{i}+x_{i+1}}{2}\right)   - g_{T}\left(\frac{x_{i-1}+x_{i}}{2}\right) \right)  - \int_{\frac{x_{i-1}+x_{i}}{2}}^{{\frac{x_{i}+x_{i+1}}{2}}} g_{T}(y) dy \\
& = J_{i,i-1} + J_{i,i+1} - \int_{\frac{x_{i-1}+x_{i}}{2}}^{{\frac{x_{i}+x_{i+1}}{2}}} g_{T}(y) dy,
\end{align*}%
and the integral in the last equality is exactly the weight of the $i$-th Voronoi cell. The expression for the density in \eqref{density} comes from the following fact: the pricing of a derivative with payoff $f$ is, recall Equation \eqref{exppol}, 
\begin{align*}
\mathbb{E}\left[f(X_{T}) \right] & = \sum_{n \geq 0} f_{n} \ell_{n}\\
& =  \sum_{n \geq 0} \int_{\mathbb{R}} f(y) H_{n}(y) w(y) dy \ \ell_{n} \\
& = \int_{\mathbb{R}} f(y) \sum_{n \geq 0} H_{n}(y)  \ell_{n}  w(y)  dy,
\end{align*}%
where the fact that we can change the order of the infinite sum and the integral is proved in  \cite{jacobi}.
Since the price of the derivative can be seen also as 
$$
\int_{\mathbb{R}} f(y) g_{T}(y) dy,
$$%
\eqref{density} follows. Finally, the expression for $\mathbb{P} \left( X_{T} \in C_{i} (\Gamma) \right)$ comes immediately from the proof of Theorem \ref{thm:masterEq}.
\end{proof}

\subsection[Proof of Lemma 4.3]{Proof of Lemma \ref{lem:error}}

\begin{proof}
First of all remember that 
$$
\textrm{err}^2_{N}  =   \left| e^{- r T } \int_{\mathbb{R}} \left( e^{x} -  K\right)^{+} g_{T}^{(M)} (x) dx  - e^{- r T } \sum_{i=1}^N \int_{C_i(\Gamma^X)} {(e^{x_i} - K)}^+ g_T^{(M)}(x) dx  \right| .
$$%
By introducing $s:= e^x$ (notice that the payoff ${(s-K)}^+$ is Lipschitz with respect to $s$ and this will be crucial), we have
\begin{eqnarray*}
	\textrm{err}^2_{N} & = & e^{- r T }  \left| \int_0^{\infty} \left( s -  K\right)^{+} \frac{g_{T}^{(M)} (\ln s) }{s} \ ds  - \sum_{i=1}^N \int_{C_i(\Gamma^S)} {( s_i - K)}^+ \frac{g_{T}^{(M)} (\ln s) }{s}  \ ds  \right|
\end{eqnarray*}%
where $S_T^{(M)}$ is a random variable with density $h_{T}^{(M)}(s) :=  \frac{g_{T}^{(M)} (\ln(s))}{s} $ for $s \in (0, + \infty)$ and where $\Gamma^S = \{ s_1, \dots, s_N \}$ is an $N$-quantizer for $S_T^{(M)}$. We denote by $\widehat S_T^{(M,N)}$ the quantization of $S_T^{(M)}$ on $\Gamma^S$. We have
$$
\mathbb{P}\left(  \widehat{S}_T^{(M,N)} = s_{i} \right) = \int_{C_{i}\left(  \Gamma^S \right)}  h_{T}^{(M)}(s) ds.
$$%
Thus, working on the error $\textrm{err}^2_{N}$ corresponds to estimating the error coming from pricing a European Call option on $S_T^{(M)}$ via quantization. Now, for every Lipschitz function $f$ with Lipschitz constant $[f]_{\text{Lip}}$, we have the following result:
\begin{align*}
\textrm{err}^2_{N} & =  \left| \mathbb{E} \left[ f\left( S_{T}^{(M)} \right) - f\left( \widehat S_T^{(M,N)} \right)  \right] \right| \\
& \leq [f]_{\text{Lip}} \left| \left|   S_{T}^{(M)} -  \widehat S_T^{(M,N)}   \right| \right| _{ 1} \\
& \leq [f]_{\text{Lip}} \left| \left|   S_{T}^{(M)} -  \widehat S_T^{(M,N)}   \right| \right| _{2},
\end{align*}%
where 
$$
|| S_T^{(M)} - \widehat{S}_T^{(M,N)} ||_{r} = \left(  \sum_{i=1}^{N} \int_{C_{i} \left(    \Gamma^S  \right)} \left|s - s_{i} \right|^{r} h_{T}^{(M)}(s) ds  \right)^{\frac{1}{r}}
$$%
is the $L^{r}$ distance between the random variable with density $h_{T}^{(M)}$ and its quantization with $N$ points. The Lipschitz constant for the payoff of a Call option is equal to one, and we have the result.
\end{proof}

\subsection[Proof of Theorem 4.4]{Proof of Theorem \ref{thm:quantError}}

\begin{proof}
It is worth noticing that in this polynomial setting, by definition (recall Equation \eqref{g_T^(M)}), the density $g_{T}^{(M)}(s), s \in (0 , + \infty),$ behaves like $s^{M}e^{-\frac{s^{2}}{2}}$, so that $h_{T}^{(M)}(s)$ behaves like $\left(\ln s \right)^{M} e^{-\frac{\left( \ln s \right)^{2}}{2}} \frac{1}{s} =: \widetilde{h}_{T}^{(M)}(s)$ at $0$ and at infinity.

The proof of \citet[Theorem 2.11]{CFG} consists of five steps, from zero to four. We now adapt it to our setting and in the case of quadratic quantization, namely in the case when $p$ in the cited paper is equal to $2$. The first three steps remain the same, so we briefly sketch them.

\smallskip\noindent
\textit{Step 0}\smallskip\\
We have to prove that $|| h_{T}^{(M)}||_{\frac{1}{p+1}}  = || h_{T}^{(M)}||_{\frac{1}{3}}  < + \infty$. We hence study the convergence at $0$ and at $+ \infty$ of the integral of $\left(\widetilde{h}_{T}^{(M)}\right)^{\frac{1}{3}}$. In the rest of the proof, without loss of generality, we will assume that $M$ is a multiple of $3$, so that computations will be explicit. If we denote by $\overline{M} := \frac{M}{3}$, then, a primitive function is 
$$
\int \left( \widetilde{h}_{T}^{(M)}(s) \right)^{\frac{1}{3}} ds =  \beta_{\overline{M}} \ Erf \left( \frac{ - 2 + \ln s}{\sqrt{6}} \right)  + \sum_{n=1}^{\overline {M}} \alpha_{n} \ s^{\frac{2}{3}} \ e^{-\frac{1}{6} \ln^{2} s } \left(\ln s \right)^{\overline{M}-n} = : \widetilde{H}^{M}_{T}(s),
$$%
where $Erf$ is the error function, defined as  $Erf(s) = \frac{2}{\pi} \int_{0}^{s} e^{-t^{2}} dt$, and the coefficients $\alpha_{n}, n=1, \dots, \overline M$ and $\beta_{\overline{M}}$ can be explicitly computed, e.g. using a symbolic programming language as \textit{Mathematica}.
Given that 
$$
\lim_{s \rightarrow +\infty} \widetilde{H}^{M}_{T}(s) = \beta_{\overline{M}}
$$%
and 
$$
\lim_{s \rightarrow 0} \widetilde{H}^{M}_{T}(s) = - \beta_{\overline{M}}
$$%
we obtain the finiteness of $||h_{T}^{(M)}||_{\frac{1}{3}}$.

\smallskip\noindent
\textit{Step 1}\smallskip\\
Here it can be shown the following estimation for the distortion function D, defined in \eqref{distortion}, associated to $S_{T}^{(M)}$ and calculated in a generic grid $\Gamma = \{ s_{1}, \dots, s_{N}\}$:
\begin{eqnarray*}
	D(s_{1}, \dots, s_{N}) & \leq & \int_{0}^{ s_{1} } (s_{1} - y)^{2}  h_{T}^{(M)}(s) ds + 
	\sum_{i=1}^{N-1} \frac{ h_{T}^{(M)}(\xi_{i})+h_{T}^{(M)}(\xi_{i+1})}{3} \left(\frac{s_{i+1} - s_{i }}{2}\right)^{3} \\ 
	& & + \int_{s_{N}}^{+ \infty } (y - s_{N})^{2}  h_{T}^{(M)}(s) ds,
\end{eqnarray*}%
for some $\xi_{1}, \dots, \xi_{N} \in \mathbb{R}$.

\smallskip\noindent
\textit{Step 2}\smallskip\\
There exists a grid $\overline{\Gamma} = \{ \overline{s}_{1}, \dots, \overline{s}_{N} \}$, and $\zeta_{1}, \dots, \zeta_{N-1}$, with     $\zeta_{i} \in [\overline{s}_{i}, \overline{s}_{i+1}]$, such that
$$
\int_{0}^{\bar{s}_{i}} \left( h_{T}^{(M)}(s)  \right)^{\frac{1}{3}}  ds  =  \int_{\bar{s}_{N}}^{+ \infty} \left( h_{T}^{(M)}(s)  \right)^{\frac{1}{3}}  ds  = \frac{1 }{2 N} || h_{T}^{(M)} ||^{\frac{1}{3}}_{\frac{1}{3}},
$$%
and
$$
\left( \overline{s}_{i+1} - \overline{s}_{i} \right)^{2}  = \frac{ || h_{T}^{(M)} ||^{\frac{2}{3}}_{\frac{1}{3}}}{ \left(h_{T}^{(M)} (\zeta_{i} ) \right)^{\frac{2}{3}} N^2}.
$$%

\smallskip\noindent
\textit{Step 3}\smallskip\\
We provide the following bound for the quantization error:
\begin{align*}
\left| \left| S_T^{(M)} - \widehat{S}_T^{(M,N)} \right| \right|^{2}_{2} & \leq \frac{ \left|\left| h_{T}^{(M)} \right|\right|^{\frac{2}{3}}_{\frac{1}{3}}}{ 24 N^2} \  \sum_{i=1}^{N-1} \frac{ h_{T}^{(M)}(\bar{\xi}_{i})+h_{T}^{(M)}(\bar{\xi}_{i+1})}{ \left(h_{T}^{(M)} (\zeta_{i} ) \right)^{\frac{2}{3}} } \left(\bar{s}_{i+1} - \bar{s}_{i }\right)  \\
& + \int_{0}^{\bar{s}_{1}} ( \bar{s}_{1} - s)^{2}  h_{T}^{(M)}(s) ds + \int_{\bar{s}_{N}}^{+ \infty } (s - \bar{s}_{N})^{2}  h_{T}^{(M)}(s) ds 
\end{align*}%

\smallskip\noindent
\textit{Step 4}\smallskip\\
In Step 2 we have proved that 
$$
\frac{1}{N^{2}} = \frac{4}{|| h_{T}^{(M)} ||^{\frac{2}{3}}_{\frac{1}{3}} } \left(\int_{0}^{\bar{s}_{i}} \left( h_{T}^{(M)}(s)  \right)^{\frac{1}{3}}  ds   \right)^{2}=  \frac{4}{|| h_{T}^{(M)} ||^{\frac{2}{3}}_{\frac{1}{3}} } \left( \int_{\bar{s}_{N}}^{+ \infty} \left( h_{T}^{(M)}(s)  \right)^{\frac{1}{3}}  ds  \right)^{2},
$$%
So in order to prove that, when $N \to \infty$, $\displaystyle  \int_{0}^{\bar{s}_{1}} ( \bar{s}_{1} - s)^{2}  h_{T}^{(M)}(s) ds$ and $\displaystyle \int_{\bar{s}_{N}}^{+ \infty } (s - \bar{s}_{N})^{2}  h_{T}^{(M)}(s) ds $ are $\displaystyle  o\left( \frac{1}{N^{2}}\right)$, we just need to prove that
$$
\lim_{{y \to +\infty}} \frac{\displaystyle   \int_{y}^{+ \infty} \left(s - y \right)^{2} h_{T}^{(M)} (s) ds}{\displaystyle \left( \int_{y}^{+ \infty}  \left( h_{T}^{(M)} (s) \right)^{\frac{1}{3}} ds \right)^{2}}  = 0
$$%
and that
$$
\lim_{{y \to 0}} \frac{  \displaystyle \int_{0}^{y} \left(s - y \right)^{2} h_{T}^{(M)} (s) ds}{\displaystyle \left( \int_{0}^{y}  \left( h_{T}^{(M)} (s) \right)^{\frac{1}{3}} ds  \right)^{2}} = 0.
$$%
Since, both at $0$ and at infinity, $h_{T}^{(M)} \sim \widetilde{h}_{T}^{(M)}$, we can equivalently prove that
$$
\lim_{{y \to +\infty}} \frac{ \displaystyle  \int_{y}^{+ \infty} s^{2} \ \widetilde{h}_{T}^{(M)} (s) ds - 2 y  \int_{y}^{+ \infty} s \ \widetilde{h}_{T}^{(M)} (s) ds + y^{2} \int_{y}^{+ \infty} \widetilde{h}_{T}^{(M)} (s) ds}{\displaystyle \left( \int_{y}^{+ \infty}  \left( \widetilde{h}_{T}^{(M)} (s) \right)^{\frac{1}{3}} ds \right)^{2}}  = 0
$$%
and that
$$
\lim_{{y \to 0}} \frac{  \displaystyle \int_{0}^{y} s^{2} \ \widetilde{h}_{T}^{(M)} (s) ds - 2 y  \int_{0}^{y} s \ \widetilde{h}_{T}^{(M)} (s) dz + y^{2} \int_{0}^{y} \widetilde{h}_{T}^{(M)} (s) ds}{ \displaystyle \left(\int_{0}^{y}  \left( \widetilde{h}_{T}^{(M)} (s) \right)^{\frac{1}{3}} ds \right)^{2}} = 0.\\
$$%
Up to a constant, we have that, for $\ell =0,1,2$, 
$$
\int s^{\ell}  \widetilde{h}_{T}^{(M)} (s) ds = \beta_{\ell,M} Erf\left( \frac{-\ell + \ln s }{\sqrt{2}} \right) + \sum_{n=0}^{M-1} \alpha_{\ell,n} e^{-\frac{1}{2} \left(\ln s \right)^{2} } s^{\ell} \left( \ln s \right)^{n} =: \widetilde{H}_{\ell,T}^{(M)} (s),
$$%
where, as before,  $Erf$ is the error function and $\alpha_{\ell,n}$ and  $\beta_{\ell,M}$ can be computed, for $\ell=0,1,2$, with \textit{Mathematica} in closed form. Please note that $\lim_{y \to +\infty} \widetilde{H}_{\ell,T}^{(M)} (y) = \beta_{\ell,M}$ and $\lim_{y \to 0} \widetilde{H}_{\ell,T}^{(M)} (y) = - \beta_{\ell,M}$, for $\ell=0,1,2$. We have then that
\begin{align*}
\lim_{{y \to +\infty}} & \frac{ \displaystyle  \int_{y}^{+ \infty} s^{2} \widetilde{h}_{T}^{(M)} (s) dz - 2 y  \int_{y}^{+ \infty} s \widetilde{h}_{T}^{(M)} (s) ds + y^{2} \int_{y}^{+ \infty} \widetilde{h}_{T}^{(M)} (s) ds}{\displaystyle \left(\int_{y}^{+ \infty}  \left( \widetilde{h}_{T}^{(M)} (s) \right)^{\frac{1}{3}} ds  \right)^{2}} = \\ 
\lim_{{y \to +\infty}}  & \frac{ \displaystyle \beta_{2,M} - \widetilde{H}_{2,T}^{(M)} (y) - 2y \left( \beta_{1,M} - \widetilde{H}_{1,T}^{(M)} (y) \right) + y^{2} \left(   \beta_{0,M} - \widetilde{H}_{0,T}^{(M)} (y)\right) }{ \displaystyle \left( \beta_{M} - \widetilde{H}_{T}^{(M)} (y) \right)^{2}} = 0,
\end{align*}%
and, in a similar way,
\begin{align*}
\lim_{{y \to 0}} & \frac{  \displaystyle \int_{0}^{y} s^{2} \widetilde{h}_{T}^{(M)} (s) ds - 2 y  \int_{0}^{y} s \widetilde{h}_{T}^{(M)} (s) ds + y^{2} \int_{0}^{y} \widetilde{h}_{T}^{(M)} (s) ds}{ \displaystyle \left( \int_{0}^{y}  \left( \widetilde{h}_{T}^{(M)} (s) \right)^{\frac{1}{3}} ds \right)^{2}} = \\  
\lim_{{y \to 0}} &  \frac{ \displaystyle \widetilde{H}_{2,T}^{(M)} (y) + \beta_{2,M} - 2y \left( \widetilde{H}_{1,T}^{(M)} (y) + \beta_{1,M}  \right) + y^{2} \left(   \widetilde{H}_{0,T}^{(M)} (y) + \beta_{0,M} \right) }{ \displaystyle  \left( \widetilde{H}_{T}^{(M)} (y) + \beta_{M} \right)^{2}} = 0.
\end{align*}%
Since
$$
\sum_{i=1}^{N-1} \frac{ h_{T}^{(M)}(\bar{\xi}_{i})+h_{T}^{(M)}(\bar{\xi}_{i+1})}{ \left(h_{T}^{(M)} (\zeta_{i} ) \right)^{\frac{2}{3}} } \left(\bar{s}_{i+1} - \bar{s}_{i }\right) \to 2 ||h_{T}^{(M)} ||_{\frac{1}{3}}^{\frac{1}{3}},
$$%
when $N \to +\infty$, we derive that 
$$
\lim_{N \to + \infty} N^{2} || S_T^{(M)} - \widehat{S}_T^{(M,N)} ||^{2}_{2} \leq \frac{1}{12} ||h_{T}^{(M)} ||_{\frac{1}{3}},
$$%
and thanks to Lemma \ref{lem:error} we conclude.
\end{proof}




\newpage

\bibliography{biblio}
\bibliographystyle{apa}

\end{document}